\documentstyle[aps,twocolumn]{revtex}

\begin{document}
\draft
\title{
Preparation and light-mediated distribution of motional state 
entanglement
}

\author{A.S. Parkins}
\address{
Department of Physics, University of Auckland, Private Bag 92019, Auckland, 
New Zealand
}
\author{E. Larsabal}
\address{
D\'epartement de Physique de
l'Ecole Normale Sup\'erieure, 24 rue Lhomond, F-75231, Paris Cedex 05, France
}

%\date{\today}

\maketitle

\begin{abstract}
We describe and analyse numerically schemes (i) for entangling orthogonal
motional modes of one or a few harmonically-trapped atoms or ions, and 
(ii) for transferring the entanglement from one of these local modes to 
a distant trapped atom (or atoms) via a light-mediated 
quantum state transfer procedure proposed in previous work 
[A.S. Parkins and H.J. Kimble, J. Opt. B: Quantum Semiclass. Opt. {\bf1}, 
496 (1999)]. Possibilities arising from these schemes include the 
generation of an Einstein-Podolsky-Rosen state in the positions and
momenta of distantly-separated trapped atoms and the preparation of 
delocalized mesoscopic vibrational states.
\end{abstract}


%\widetext

\section{
Introduction
}

In addition to being of fundamental interest in physics, 
quantum entanglement is an essential resource for quantum information 
processing and distribution. Popular candidates for experimental
investigation in this context include trapped atoms, cavity quantum
electrodynamics (cavity QED), and nonclassical 
light fields, using which a variety of impressive quantum state 
manipulations have now been demonstrated, including elementary quantum
logic operations \cite{Brune94,Turchette95,Monroe95,Rauschenbeutel99}, 
preparation of quantum-mechanically entangled pairs 
\cite{Hagley97,Turchette98,Rauschenbeutel00} and quadruplets 
\cite{Sackett00} of atoms, and quantum 
teleportation \cite{Bouwmeester97,Boschi98,Furusawa98}. 

In recent work we have proposed a scheme that offers the possibility
of combining the features of trapped atoms and nonclassical light
fields in a distributed network \cite{Parkins99};
in particular, a scheme that enables motional quantum states to be
coupled to propagating (nonclassical) light fields via interactions
in cavity QED.
This coupling enables the {\em deterministic} 
generation and distribution of quantum
entanglement between different atoms and/or different light fields.
One particular example we have considered is the preparation of an 
Einstein-Podolsky-Rosen (EPR) state in the positions and momenta of
a pair of distantly-separated trapped atoms \cite{Parkins00a}, 
which in turn leads to a scheme for the teleportation
of motional quantum states \cite{Parkins00b}.

For the proposals presented in \cite{Parkins00a,Parkins00b} 
entanglement of the atoms' motional states is achieved through the
transfer of entanglement from the quantum-correlated output light 
fields from a nondegenerate parametric amplifier. 
However, as pointed out in \cite{Parkins99}, the trapped-atom cavity 
QED setup can itself act as a source of nonclassical light fields;
nonclassical motional states (prepared by some means independent of 
the atom-cavity coupling) can be transferred to the propagating 
output field from the cavity, which can in turn be coupled to the
motion of a second atom confined inside another, distant cavity.
The potential of such a trapped-atom nonclassical light source is
underscored by the impressive control with which the motional quantum
states of single trapped atoms can be controlled and, more especially,
by the variety of nonclassical motional states that have in fact 
been prepared experimentally 
\cite{Meekhof96,Monroe96,Wineland98a,Roos99}.

The purpose of the present paper, then, is to expand upon this idea
and present schemes whereby ``local'' entanglement of orthogonal motional
modes of a single atom (or of a few atoms) trapped inside an optical 
cavity is transformed, via propagating light fields, into ``nonlocal''
entanglement of the motional modes of distantly-separated atoms. In this
way, the need for nonlinear (quantum) optical devices, such as 
nondegenerate parametric amplifiers, is eliminated, and all of the desired
operations are achieved using only trapped-atom cavity QED configuratons.

In Section II we present a relatively simple example of a scheme for the
manipulation of the motional state of a trapped atom (or atoms) in two
dimensions using only (external) laser fields. This scheme allows 
entanglement to be established between two orthogonal modes and the
results of some numerical simulations
for the case of a two-mode ``squeezed'' state of the motion are presented. 
Section III describes the trapped-atom cavity QED system that facilitates
the motion-light coupling and is, of course, central to our proposals;
through this coupling, the quantum state (i.e., the entanglement) of one 
of the two relevant motional modes can be ``mapped onto'' a propagating 
light field and transferred to a distant site. We examine this system in 
some detail, presenting numerical calculations that further the original 
analysis of \cite{Parkins99}, looking more closely, for example, at the
Lamb-Dicke (tight confinement) assumption involved in the model.
Having described means for preparing ``local'' entanglement of motional modes
and for transferring motional quantum states between distant atoms, the
preparation of distributed ``nonlocal'' motional state entanglement follows
naturally, and in Section IV we discuss several examples, such as the
position-momentum EPR state and delocalized mesoscopic states of the motion.

\section{
Entangling orthogonal motional modes at one location
}

Schemes for coupling and manipulating orthogonal motional modes in a 
single-atom trap have been put forward by a variety of authors 
\cite{Gou96a,Gou96b,Steinbach97,Gardiner97,Drobny97,Drobny98,Wineland98b,Kneer98}, 
typically involving stimulated-Raman transitions between internal 
atomic states.  
Here, we describe one such scheme, 
outlining how certain effective interaction Hamiltonians 
(of particular interest to us) may be
realized and presenting some numerical simulations. 
This scheme involves only a 
single internal atomic level (after adiabatic elimination of the excited
internal state) and generalizes to two dimensions a 
technique already used in ion-trap experiments to prepare, for example,
one-dimensional squeezed states of the motion \cite{Meekhof96}.

\subsection{
Coupling orthogonal motional modes of a single trapped atom
}

In this scheme, counterpropagating laser beams aligned (in our case)
in the $x$-$z$ plane and coupled to the same internal atomic transition
induce a coupling between orthogonal ($x$ and $z$)
motional modes which, depending on 
the detuning between the two laser fields, may take the form of a linear
mixer or a nondegenerate parametric amplifier. 

The physical setup is depicted in Fig.~1(a).
Mathematically, the situation is described by the Hamiltonian
\begin{eqnarray} \label{2modeH}
\hat{H}(t) = && 
\hbar\Delta_{01}\hat{\sigma}_+\hat{\sigma}_- +
\sum_{j=x,z} 
\hbar\nu_j (\hat{b}_j^\dagger\hat{b}_j+1/2) 
\nonumber
\\
&& \; +\, i\hbar 
\left[ E_{\rm L}^\ast (\hat{x},\hat{z},t) 
\hat{\sigma}_- - 
E_{\rm L} (\hat{x},\hat{z},t)
\hat{\sigma}_+ \right] ,
\end{eqnarray}
where $\nu_x$ and $\nu_z$ are the harmonic oscillation frequencies along
the $x$ and $z$ axes of the trap, $\hat{b}_j$ are annihilation
operators for the quantized atomic motion,
and $\hat{\sigma}_-=|g\rangle\langle e|$ is the atomic lowering operator
for the $|g\rangle\leftrightarrow |e\rangle$ transition.
The total field incident on the atom is given by
\begin{eqnarray} \label{E_L}
E_{\rm L} (\hat{x},\hat{z},t) &=&
E_1(\hat{x},\hat{z},t) + E_2(\hat{x},\hat{z},t)
\nonumber
\\
&=&
\frac{{\cal E}}{\sqrt{2}} 
\left[ \, e^{-ik(\alpha\hat{x}+\beta\hat{z})}
+ e^{ik(\alpha\hat{x}+\beta\hat{z})-i\delta_{21} t+i\phi} \, \right] ,
\end{eqnarray}
where $\alpha$ and $\beta$ are determined by the angle of incidence
in the $x$-$z$ plane and satisfy $(\alpha^2+\beta^2)^{1/2}=1$.
The position operators are given by
$\hat{x}=[\hbar /(2m\nu_x)]^{1/2}(\hat{b}_x+\hat{b}_x^\dagger )$ and
$\hat{z}=[\hbar /(2m\nu_z)]^{1/2}(\hat{b}_z+\hat{b}_z^\dagger )$, and
$k=2\pi /\lambda$ is the wavenumber (taken to be the same for both fields).
The detunings $\Delta_{01}$ and $\delta_{21}$ are given by
$\Delta_{01} =\omega_0-\omega_1$ and
$\delta_{21} =\omega_2-\omega_1$, where $\omega_0$ is the atomic
transition frequency and $\omega_1$ and $\omega_2$ are
the frequencies of the two counterpropagating fields. 

With the 
assumption that $\Delta_{01}$ is large, the atomic excited state
can be adiabatically eliminated and spontaneous emission neglected.
In the Heisenberg equations of motion one makes the substitution
\begin{equation}
\hat{\sigma}_- \simeq i\, 
\frac{E_{\rm L} (\hat{x},\hat{z},t)}{\Delta_{01}} ,
\end{equation}
which gives, for the motional mode operators,
\begin{eqnarray}
\dot{\hat{b}}_{x,z} \simeq && -i\nu_{x,z}\hat{b}_{x,z}
+ \frac{2i}{\Delta_{01}} \, \left[ \hat{b}_{x,z},
|E_{\rm L} (\hat{x},\hat{z},t)|^2 \right]
\\
=&& -i\nu_{x,z}\hat{b}_{x,z} \nonumber
\\
&& +
\frac{2\eta_{x,z}^\prime {\cal E}^2}{\Delta_{01}} \, \left[
e^{-2ik(\alpha\hat{x}+\beta\hat{z})+i\delta_{21}t-i\phi} -
{\rm h.c.} \right] ,
\end{eqnarray}
where $\eta_x^\prime =\alpha\eta_x$ and $\eta_z^\prime =\beta\eta_z$,
with $\eta_x=k(\hbar /2m\nu_x)^{1/2}$ and 
$\eta_z=k(\hbar /2m\nu_z)^{1/2}$.

Moving to a rotating frame with respect to the motion, i.e.,
defining $\hat{b}_{x,z}=\tilde{b}_{x,z}e^{-i\nu_{x,z}t}$,
and making the Lamb-Dicke approximation, i.e., expanding the exponentials
to first order in $\eta_{x,z}^\prime$, one derives
\begin{eqnarray} \label{tbxd}
\dot{\tilde{b}}_x \simeq &&
\frac{2\eta_x^\prime{\cal E}^2}{\Delta_{01}} \,
e^{i\nu_xt} \left( e^{i\delta_{21} t-i\phi} - e^{-i\delta_{21}t+i\phi}
\right) \nonumber
\\
&& -\, \frac{4i{\eta_x^\prime}^2{\cal E}^2e^{-i\phi}}{\Delta_{01}} \,
e^{i(\nu_x+\delta_{21})t} \left( \tilde{b}_x
e^{-i\nu_xt} + \tilde{b}_x^\dagger 
e^{i\nu_xt} \right) \nonumber
\\
&& -\, \frac{4i\eta_x^\prime\eta_z^\prime {\cal E}^2e^{-i\phi}}
{\Delta_{01}} \,
e^{i(\nu_x+\delta_{21})t} \left( \tilde{b}_z
e^{-i\nu_zt} + \tilde{b}_z^\dagger 
e^{i\nu_zt} \right) \nonumber
\\
&& -\, \frac{4i{\eta_x^\prime}^2{\cal E}^2e^{i\phi}}{\Delta_{01}} \,
e^{i(\nu_x-\delta_{21})t} \left( \tilde{b}_x
e^{-i\nu_xt} + \tilde{b}_x^\dagger 
e^{i\nu_xt} \right) \nonumber
\\
&& -\, \frac{4i\eta_x^\prime\eta_z^\prime{\cal E}^2e^{i\phi}}
{\Delta_{01}} \,
e^{i(\nu_x-\delta_{21})t} \left( \tilde{b}_z
e^{-i\nu_zt} + \tilde{b}_z^\dagger 
e^{i\nu_zt} \right) 
\end{eqnarray}
(and similarly for $\dot{\tilde{b}}_z$). This equation contains a
variety of terms oscillating at a variety of different frequencies.
With a judicious choice of the detuning $\delta_{21}$ between the two
laser fields, a particular term can have its time dependence removed,
whereas the other (oscillating) terms can, provided the trap frequencies
are sufficiently large (and {\em different}), 
be neglected in a rotating-wave approximation. 
We now consider two cases of interest to us.

\subsubsection{
Linear mixing: rotation of the motional state
}

To realize a linear mixing of the $x$ and $z$ modes (i.e., a
beamsplitter-type interaction), we choose the detuning between the
two incident laser fields to be $\delta_{21}=\nu_x-\nu_z$. With this
choice, only one term in (\ref{tbxd}) is ``stationary'', while all of the
other terms retain oscillatory factors. Provided the frequencies
$\nu_x$, $\nu_z$, and $|\nu_x-\nu_z|$ are large compared to the
effective coupling strength between the two modes, these oscillatory
terms can be dropped in a rotating-wave approximation, leaving
\begin{equation}
\dot{\tilde{b}}_{x,z} \simeq - \, i\left(
\frac{4\eta_x^\prime\eta_z^\prime{\cal E}^2e^{\pm i\phi}}
{\Delta_{01}} \right) \, \tilde{b}_{z,x} .
\end{equation}
The effective interaction Hamiltonian is thus
\begin{equation} \label{mixer}
\tilde{H}_{\rm mix} = \hbar \chi 
\left( \tilde{b}_x^\dagger \tilde{b}_z e^{i\phi}
+ {\rm h.c.} \right) ,
\end{equation}
with the interaction strength given by
\begin{equation} \label{chi}
\chi = 
\frac{4\eta_x^\prime\eta_z^\prime {\cal E}^2}{\Delta_{01}} .
\end{equation}
The physical process corresponding to this situation is illustrated
in Fig.~1(b). 
Particular cases of interest are those in which the interaction is
turned on for a time $T$ such that $\chi T=\pi /4$ or $\chi T=\pi /2$
(we assume rectangular pulse shapes); the first case effects an operation 
equivalent to a 50/50 beamsplitter, while in the second case the result
is a complete exchange of states between the two modes, i.e.,
for $\phi =\pi /2$ one finds 
$\tilde{b}_x(T)=-\tilde{b}_z(0)$ and $\tilde{b}_z(T)=\tilde{b}_x(0)$.

Note also that, given suitable (unentangled) initial states of the $x$ and
$z$ modes (prepared by some other means), one can generate entanglement
with the beamsplitter ($\chi T=\pi /4$) operation. For example, an initial
product of number states,
\begin{equation}
|\psi (0)\rangle = |N_x\rangle_x\otimes |N_z\rangle_z ,
\end{equation}
is transformed via a 50/50 beamsplitter operation into an entangled state
of the form 
\begin{equation}
|\psi (T)\rangle = \sum_{n=0}^N d_n |N-n\rangle_x\otimes |n\rangle_z ,
\end{equation}
where $N=N_x+N_z$ and $\{ d_n\}$ are certain coefficients.

\subsubsection{
Two-mode squeezing of the motional state
}

Alternatively, we may choose $\delta_{21}=\nu_x+\nu_z$, in which case, 
following the same assumptions as above, one derives
\begin{equation}
\dot{\tilde{b}}_{x,z} \simeq - \, i\left(
\frac{4\eta_x^\prime\eta_z^\prime{\cal E}^2e^{\pm i\phi}}
{\Delta_{01}} \right) \,
\tilde{b}_{z,x}^\dagger , 
\end{equation}
corresponding to
\begin{equation}
\tilde{H}_{\rm sq} = \hbar \chi 
\left( \tilde{b}_x^\dagger \tilde{b}_z^\dagger e^{i\phi}
+ {\rm h.c.} \right) ,
\end{equation}
with $\chi$ as before. The physical process corresponding to this
situation is illustrated in Fig.~1(c); it is of course a
parametric amplification process, which leads to ``two-mode
squeezing''. Taking $\phi =-\pi /2$, the state produced after an
interaction time $T$, given the modes are initially in their ground
vibrational levels (achieved, for example, by sideband cooling), is
\begin{eqnarray} \label{psixz}
|\psi_{\rm sq}(r)\rangle &=& S_{xz}(r) |0\rangle_{x}\otimes |0\rangle_{z} 
\nonumber
\\
&=& \left[ \cosh (r) \right]^{-1} \sum_{m=0}^\infty
\left[ -\tanh (r) \right]^m \, |m\rangle_{x}\otimes |m\rangle_{z} \, ,
\end{eqnarray}
where $|m\rangle_{x,z}$ are Fock states of the motional
modes and $S_{xz}(r)$ is the two-mode squeezing operator
\cite{Walls94},
\begin{equation}
S_{xz}(r) = \exp \left[ r\left( \tilde{b}_{x}\tilde{b}_{z} -
\tilde{b}_{x}^\dagger\tilde{b}_{z}^\dagger 
\right) \right] \, ,
\end{equation}
with $r=\chi T$.
The entanglement between modes that is generated by this process
is best expressed in terms of position and momentum variables.
In particular, one can show for the state (\ref{psixz}) that
the Wigner function in these variables is
\begin{eqnarray}
&& W(\tilde{x},\tilde{p}_x;\tilde{z},\tilde{p}_z) \nonumber
\\
&& \;\;\;\;\;\;\; = \frac{4}{\pi^2} \, 
\exp \left\{ -\left[ (\tilde{x}+\tilde{z})^2+
(\tilde{p}_x-\tilde{p}_z)^2\right] e^{+2r} \right\} \nonumber
\\
&& \;\;\;\;\;\;\;\;\;\;\;\;\;\; \times \; 
\exp \left\{ -\left[ (\tilde{x}-\tilde{z})^2+
(\tilde{p}_x+\tilde{p}_z)^2\right] e^{-2r} \right\} 
\\
&& \;\;\;\;\;\;\; \sim
C\; \delta (\tilde{x}+\tilde{z})\,\delta (\tilde{p}_x-\tilde{p}_z) \;\;\;
{\rm for} \;{\rm large}\;\, r ,
\end{eqnarray}
with $C$ a constant \cite{Walls94,Ou92b}. 
This, of course, corresponds to the original
state considered by Einstein, Podolsky, and Rosen in their famous
gedanken experiment \cite{Einstein35}, although, in this particular 
situation the position and momentum variables belong to the same
particle. However, as we will describe below, 
through our scheme for motion-light coupling we
are able to distribute this entanglement between two 
distantly-separated atoms.

\subsection{
Numerical analysis
}

Before describing the scheme for distribution of motional state
entanglement, we consider the results of some numerical simulations
of the state manipulation schemes proposed above.
By numerical simulation, we mean direct integration of the
Schr\"odinger equation
\begin{equation}
\frac{\partial}{\partial t}
|\psi (t)\rangle = \frac{1}{i\hbar} \hat{H}_{\rm ad}(t)
|\psi (t)\rangle ,
\end{equation}
with $\hat{H}_{\rm ad}(t)$ given by (omitting the constant ground state
vibrational energy)
\begin{equation}
\hat{H}_{\rm ad}(t) = \hbar\nu_x\hat{b}_x^\dagger\hat{b}_x +
\hbar\nu_z\hat{b}_z^\dagger\hat{b}_z - \hbar\,
\frac{2|E_{\rm L}(\hat{x},\hat{z},t)|^2}{\Delta_{01}} ,
\end{equation}
which is the form one obtains upon adiabatically eliminating the atomic
excited state.
Truncated number state bases are used to describe the harmonic oscillator 
modes describing the two-dimensional motion. 

We have focussed in our work on the two-mode squeezing, or 
EPR entanglement, operation of the previous subsection,
determining in particular suitable parameter regimes for the
efficient implementation of the scheme 
(requirements for the linear mixing scheme are basically the same).
As a measure of how well the scheme works, we consider the
fidelity
\begin{equation}
F = |\langle\psi_{\rm sq}(r)|\psi (T)\rangle_{\rm sim} |^2 ,
\end{equation}
where $|\psi (T)\rangle_{\rm sim}$ is the simulated wave function after
a time $T$, and $r=\chi T$ with $\chi$ given by (\ref{chi}).

\paragraph{Results}

Sample values of $F$ are given in Table I for a number of different
combinations of trapping frequencies and Lamb-Dicke parameters. 
As one can see, two-mode squeezed states exhibiting significant degrees of
entanglement ($r\gtrsim 1$) can be prepared with high fidelity given
sufficiently small Lamb-Dicke parameters and sufficiently large
$\nu_x$, $\nu_z$, and $|\nu_x-\nu_z|$ (compared to $\chi$).

It becomes computationally prohibitive for us to consider values of
$r$ much larger than $\sim 1.5$, owing to the large number state bases
required to cover adequately the population distribution of the two-mode
squeezed state. However, the results presented here suggest that, for example,
a state with $r\simeq 2$ could be prepared with a fidelity exceeding
0.9 for physically reasonable parameters. 

\paragraph{Experimental prospects}

If we take the example of trapped ${}^9{\rm Be}^+$ ions, 
$\eta_x^\prime =\eta_z^\prime =0.0707$ and $\nu_x/\nu_z=1/3$ correspond
to actual trap frequencies $\nu_x/2\pi =11\;{\rm MHz}$ and
$\nu_z/2\pi =34\;{\rm MHz}$ (which are essentially the frequencies achieved
in the experiments of Monroe and coworkers \cite{Monroe96}), 
and thus, for the example given in the table,
to a value of the interaction strength $\chi /2\pi =22\;{\rm kHz}$ and
a timescale for the preparation of the entangled state on the order of 
$\chi^{-1}\simeq 7\;\mu {\rm s}$. This should of course be much smaller 
than the timescale for motional state decoherence due to, for example,
spontaneous emission, which we have neglected in our analysis. The rate
of spontaneous emission events influencing the motional dynamics
(i.e., incoherent scattering) can be estimated to be on the order of
$(\gamma\eta_x^2{\cal E}^2)/\Delta_{01}^2$, 
where $\gamma^{-1}$ is the atomic excited state lifetime (see Appendix A,
Part 1). 
Choosing $\gamma /\Delta_{01}\ll 1$, this rate can evidently be made 
much smaller than $\chi$.

\subsection{
Two or more trapped ions: coupling collective and single ion modes
}

Two or more {\em ions} confined in a linear ion trap \cite{Raizen92}
interact strongly
through Coulomb repulsion and their motion along the axis of the trap
is best described in terms of collective modes of vibration. 
Such collective modes have, of course, been fundamental to ion-trap
quantum computer proposals \cite{Cirac95a}.
In the present context, we wish to point out that the schemes of
Section II.A could also be used to couple and entangle orthogonal 
collective modes of a linear chain of trapped ions.

To illustrate this, we consider for simplicity just a pair of trapped
ions confined in a harmonic potential and aligned along the $z$-axis,
as illustrated in Fig.~2. Assuming very strong confinement in the $x$
direction (so that the motion along this axis can be regarded as 
independent of the $z$-axis collective motion), the quantized
(two-dimensional) motion of the ions can be described by the
Hamiltonian
\begin{eqnarray}
\hat{H}_{\rm mot} && = \hbar\nu_x\sum_{j=1}^2 \left( \hat{b}_{jx}^\dagger
\hat{b}_{jx} + \frac{1}{2} \right) \nonumber
\\
&& +\, \hbar\nu_{0z}\left( 
\hat{c}_{0z}^\dagger \hat{c}_{0z} + \frac{1}{2} \right)
+ \hbar\nu_{Rz}\left( 
\hat{c}_{Rz}^\dagger \hat{c}_{Rz} + \frac{1}{2} \right) ,
\end{eqnarray}
where $\hat{b}_{jx}$, $\hat{c}_{0z}$, and $\hat{c}_{Rz}$ are harmonic
oscillator annihilation operators for single-ion motion along the
$x$-axis, for the center-of-mass motion along the $z$-axis, and for 
the relative motion along the $z$-axis, respectively.
The frequencies of the collective modes are related to the single-ion
mode frequency by $\nu_{0z}=\nu_z$ and $\nu_{Rz}=\sqrt{3}\nu_z$
\cite{Cirac95a,James98}.

To couple the various modes via laser light, excitation of only a single
ion is necessary, and this excitation takes the same form as described by
(\ref{2modeH}) and (\ref{E_L}); that is, we take
\begin{eqnarray}
\hat{H}_{\rm ion-laser} && = 
\hbar\Delta_{01}\hat{\sigma}_+^{(1)}\hat{\sigma}_-^{(1)} \nonumber
\\
&& + i\hbar 
\left[ E_{\rm L}^\ast (\hat{x}_1,\hat{z}_1,t) 
\hat{\sigma}_-^{(1)} - {\rm h.c.}
\right] ,
\end{eqnarray}
with
\begin{eqnarray}
E_{\rm L} &&(\hat{x}_1,\hat{z}_1,t) = \nonumber
\\
&&
\frac{{\cal E}}{\sqrt{2}} 
\left[ \, e^{-ik(\alpha\hat{x}_1+\beta\hat{z}_1)}
+ e^{ik(\alpha\hat{x}_1+\beta\hat{z}_1)-i\delta_{21}t+i\phi} \, \right] .
\end{eqnarray}
Expressed in terms of the collective mode operators, one has
\cite{Morigi99}
\begin{equation}
e^{ik\hat{z}_1} = e^{i\eta_{0z}(\hat{c}_{0z}^\dagger+\hat{c}_{0z})}
e^{i\eta_{Rz}(\hat{c}_{Rz}^\dagger+\hat{c}_{Rz})} ,
\end{equation}
with the collective mode Lamb-Dicke parameters given by
\begin{equation}
\eta_{0z} = \frac{\eta_z}{\sqrt{2}} , \;\;\;\;
\eta_{Rz} = \frac{\eta_z}{\sqrt{2\sqrt{3}}} .
\end{equation}
As before, the atomic excited state is adiabatically eliminated and the
position-dependent functions are expanded to first order in the various 
Lamb-Dicke parameters. The resulting Heisenberg equations of motion for
the mode operators again display couplings between the modes with varying
time dependencies, and again with a judicious choice of the detuning 
$\delta_{21}$ one can ``select'' a particular coupling (neglecting the others
in a rotating-wave approximation), provided the frequencies 
$\nu_x$, $\nu_{0z}$, and $\nu_{Rz}$ are sufficiently large and different.
The effective interaction Hamiltonians one can realize are thus
(in the rotating frame)
\begin{equation}
\tilde{H}_{\rm eff} = \left\{  
\begin{array}{ll}
\hbar\chi_0 \left( \tilde{b}_{1x}^\dagger \tilde{c}_{0z} e^{i\phi}
+ {\rm h.c.} \right) , & \delta_{21} =\nu_x-\nu_{0z} \\
\hbar\chi_0 \left( \tilde{b}_{1x}^\dagger \tilde{c}_{0z}^\dagger e^{i\phi}
+ {\rm h.c.} \right) , & \delta_{21} =\nu_x+\nu_{0z} \\
\hbar\chi_R \left( \tilde{b}_{1x}^\dagger \tilde{c}_{Rz} e^{i\phi} 
+ {\rm h.c.} \right) , & \delta_{21} =\nu_x-\nu_{Rz} \\
\hbar\chi_R \left( \tilde{b}_{1x}^\dagger \tilde{c}_{Rz}^\dagger e^{i\phi} 
+ {\rm h.c.} \right) , & \delta_{21} =\nu_x+\nu_{Rz} 
\end{array}
\right.
\end{equation}
with
\begin{equation}
\chi_0 = \frac{4\eta_x^\prime\eta_{0z}^\prime {\cal E}^2}{\Delta_{01}} ,
\;\;\;\;
\chi_R = \frac{4\eta_x^\prime\eta_{Rz}^\prime {\cal E}^2}{\Delta_{01}} ,
\end{equation}
and $\eta_x^\prime =\alpha\eta_x$, $\eta_{0z}^\prime =\beta\eta_{0z}$,
$\eta_{Rz}^\prime =\beta\eta_{Rz}$.
So, one again realizes linear mixers and parametric amplifiers, only
now one of the modes is a collective mode for the two ions.
With the addition of more ions, one introduces more collective modes,
but the frequencies of these modes remain incommensurate and (for large
$\nu_z$) well-separated \cite{James98}, implying that the above
working could be readily generalized to three or more ions.

\section{
Coupling motion to light
}

\subsection{
Cavity-mediated light-motion coupling
}

\subsubsection{
Model
}

The basic setup we use to couple motion to light
was originally considered by Zeng and Lin \cite{Zeng94}. 
This setup consists of a two-level atom confined
in a harmonic trap located inside an optical cavity. 
The atomic transition of frequency
$\omega_0$ is coupled to a single mode of the cavity field of
frequency $\omega_{\rm cav}$ and is also driven by an
external (classical) laser field of frequency $\omega_{\rm A}$.
The physical setup and excitation scheme are
depicted in Fig.~3.
The cavity is aligned along the $x$-axis, while
the laser field is incident from a direction 
along the $y$-axis (i.e., perpendicular to the $x$-axis).

The Hamiltonian describing the atom-cavity system, including the
atomic motion, takes the  
form (in a frame rotating at the laser frequency, $\omega_{\rm A}$)
\begin{eqnarray} \label{H}
\hat{H}_{\rm ac} = && 
\sum_{j=x,y,z} 
\hbar\nu_j (\hat{b}_j^\dagger\hat{b}_j+1/2) + \hbar\delta_{\rm cA}
\hat{a}^\dagger\hat{a} + \hbar\Delta_{\rm 0A}\hat{\sigma}_+\hat{\sigma}_-
\nonumber
\\
&& \;\;\; +\, \hbar 
\left[ E_{\rm A}(\hat{y},t) 
\hat{\sigma}_+ + 
E_{\rm A}^\ast (\hat{y},t)
\hat{\sigma}_- \right] \nonumber
\\
&& \;\;\; +\, \hbar
g_0 \sin (k\hat{x}) (\hat{a}^\dagger
\hat{\sigma}_- + \hat{\sigma}_+\hat{a} )  \nonumber
\\
&& \;\;\; +\, 
\hat{a}^\dagger \hat{\Upsilon}_{\rm c} 
+ \hat{\Upsilon}_{\rm c}^\dagger \hat{a} 
+ \hat{\sigma}_+ \hat{\Upsilon}_{\rm a} 
+ \hat{\Upsilon}_{\rm a}^\dagger \hat{\sigma}_- .
\end{eqnarray}
Here, $\{\nu_x,\nu_y,\nu_z\}$ are the harmonic oscillation 
frequencies along the principal axes of the trap, 
$\hat{b}_j$ and $\hat{a}$ are annihilation operators for the
quantized atomic motion and cavity field, respectively, 
$\hat{\sigma}_-=|g\rangle\langle e|$ is the atomic lowering 
operator, and $\delta_{\rm cA} =\omega_{\rm cav}-\omega_{\rm A}$ and 
$\Delta_{\rm 0A} =\omega_0-\omega_{\rm A}$. 
The quantity $E_{\rm A}(\hat{y},t)$ is the
(possibly time-dependent) amplitude of laser field A.
The single-photon atom-cavity dipole coupling strength is
given by $g_0$, while the sine function describes the standing wave
structure of the cavity field -- we assume that the centre of the 
trap is located at a {\em node} of the cavity field. 
Finally, the last two terms in (\ref{H}) describe the couplings of
the cavity field mode and the atomic transition to ``reservoirs'' 
of external field modes (with $\hat{\Upsilon}_{\rm a,c}$ the ``reservoir
annihilation operators''), which produce damping of the cavity field
and (free-space) atomic spontaneous emission, respectively 
\cite{Walls94}. 
Note that we will neglect any forms of motional decoherence 
associated with the trap itself.

\subsubsection{
Elimination of the atomic excited state
}

Heisenberg equations of motion are straightforwardly derived from
the above Hamiltonian. Assuming the detunings of the light fields from 
the atomic transition frequency to be very large 
[i.e., $\Delta_{\rm 0A}\gg |E_{\rm A}|,g_0,\delta,\nu_j$],
atomic spontaneous emission can be neglected and the internal atomic
dynamics can be adiabatically eliminated. In the equations of motion,
this is done by making the replacement
\begin{equation}
\hat{\sigma}_- \simeq -\frac{1}{\Delta_{\rm 0A}} \left[ 
E_{\rm A}(\hat{y},t) + g_0\sin (k\hat{x}) \hat{a}
\right] 
\end{equation}
in the equations describing the cavity and motional degrees of
freedom. 
The corresponding Hamiltonian then takes the form
\begin{eqnarray} \label{Hac}
\hat{H}_{\rm ac} = && 
\sum_{j=x,y,z} 
\hbar\nu_j (\hat{b}_j^\dagger\hat{b}_j+1/2) + \hbar\delta_{\rm cA}
\hat{a}^\dagger\hat{a} \nonumber
\\
&& \;\;\; -\, \frac{\hbar |E_{\rm A}(\hat{y},t)|^2}
{\Delta_{\rm 0A}} - \frac{\hbar g_0^2}{\Delta_{\rm 0A}} \, \sin^2(k\hat{x}) 
\hat{a}^\dagger\hat{a}
\nonumber
\\
&& \;\;\; -\, \frac{\hbar g_0}{\Delta_{\rm 0A}}\, 
\sin (k\hat{x}) \left[
E_{\rm A}(\hat{y},t)\hat{a}^\dagger
+ E_{\rm A}^\ast (\hat{y},t)\hat{a} \right]  \nonumber
\\
&& \;\;\; +\, 
\hat{a}^\dagger \hat{\Upsilon}_{\rm c} 
+ \hat{\Upsilon}_{\rm c}^\dagger \hat{a} .
\end{eqnarray}

\subsubsection{
Lamb-Dicke and rotating-wave approximations
}

The size of the harmonic trap is assumed to be small compared to the
optical wavelength (Lamb-Dicke approximation); this 
enables the approximations
$\sin (k\hat{x})\simeq \eta_x (\hat{b}_x+\hat{b}_x^\dagger )$, and 
$E_{\rm A}(\hat{y},t)\simeq
{\cal E}_{\rm A}(t){\rm e}^{-i\phi_{\rm A}}$ [with
${\cal E}_{\rm A}(t)$ real]. This second approximation
would follow, for example, if the laser field forms a standing wave
with the trap centered at an antinode, i.e., with 
$E_{\rm A}(\hat{y})\propto\cos (k\hat{y})\simeq 1$, for
$\eta_y\ll 1$ (for further discussion of this approximation, see
Appendix B).

To second order in $\eta_x$, quantum Langevin equations for the
field and $x$-dimension motional modes are then
\begin{eqnarray}
\dot{\hat{a}} &=& -(\kappa +i\delta_{\rm cA})\hat{a} 
+ i\frac{\eta_x^2g_0^2}{\Delta_{\rm 0A}} \left( \hat{b}_x+
\hat{b}_x^\dagger \right)^2\hat{a}
\nonumber
\\
&& \; -\, i\Omega (t)
e^{-i\phi_{\rm A}} \left( \hat{b}_x+\hat{b}_x^\dagger \right) 
 -\, \sqrt{2\kappa}\,e^{-i\delta_{\rm cA}t}
\hat{a}_{\rm in}(t) , 
\\
\dot{\hat{b}}_x &=& -i\nu_x\hat{b}_x 
+ i\frac{2\eta_x^2g_0^2}{\Delta_{\rm 0A}} \left( \hat{b}_x+
\hat{b}_x^\dagger \right) \hat{a}^\dagger\hat{a}
\nonumber
\\
&& \; -\, i\Omega (t) \left(
\hat{a}^\dagger e^{-i\phi_{\rm A}} + \hat{a} e^{i\phi_{\rm A}} 
\right) ,
\end{eqnarray}
where we have defined 
\begin{equation}
\Omega (t) = - \frac{\eta_xg_0{\cal E}_{\rm A}(t)}{\Delta_{\rm 0A}} ,
\end{equation}
while the operator $\hat{a}_{\rm in}(t)$ obeys the commutation
relation 
$[\hat{a}_{\rm in}(t),\hat{a}_{\rm in}^\dagger (t^\prime )]
=\delta (t-t^\prime )$ and describes the quantum noise input to
the cavity field from the external field (in a frame rotating at the
cavity frequency). The parameter $\kappa$ is the cavity field decay
rate.

Next, we choose the detuning between the cavity and laser fields 
to be $\delta_{\rm cA}=\omega_{\rm cav}-\omega_{\rm A}=\nu_x$. In the 
above equations this results in ``resonant'' and ``non-resonant'' terms.
With the assumption that the trap frequency $\nu_x$ is large
(which is consistent with the Lamb-Dicke assumption, 
since $\eta_x\propto\nu_x^{-1/2}$), such that
$\nu_x\gg\kappa ,|\Omega(t)|$, the non-resonant, or counter-rotating
terms can be neglected in a rotating-wave approximation.
This leads to the pair of equations
\begin{eqnarray}
\dot{\hat{a}} &=& -(\kappa +i\nu_x )\hat{a} 
+ i\frac{2\eta_x^2g_0^2}{\Delta_{\rm 0A}} \left( 
\hat{b}_x^\dagger\hat{b}_x + \frac{1}{2} \right) \hat{a}
\nonumber
\\
&& \; -\, i\Omega (t)
e^{-i\phi_{\rm A}} \hat{b}_x
 -\, \sqrt{2\kappa}\, e^{-i\nu_x t}
\hat{a}_{\rm in}(t) , 
\\
\dot{\hat{b}}_x &=& -i\nu_x\hat{b}_x 
+ i\frac{2\eta_x^2g_0^2}{\Delta_{\rm 0A}}
\hat{a}^\dagger\hat{a} \hat{b}_x
- i\Omega (t) e^{i\phi_{\rm A}} \hat{a} .
\end{eqnarray}
The terms of second order in $\eta_x$ describe phonon- or
photon-number-dependent phase shifts; these will in general be
very small and can be neglected (although they are retained for
numerical calculations) \cite{Harrison97}, 
which means that the effective interaction between the 
cavity and motional modes is simply a linear coupling of the form
\begin{equation}
\hat{H}_{\rm eff} = \hbar\Omega (t) \left( \hat{a}^\dagger
\hat{b}_x e^{-i\phi_{\rm A}} + 
\hat{b}_x^\dagger \hat{a} e^{i\phi_{\rm A}} \right) .
\end{equation}

\subsubsection{
Adiabatic elimination of the cavity mode
}

Although not essential for our purposes, a further
simplification of the dynamics is possible if the decay rate 
$\kappa$ of the cavity field is sufficiently large that the
cavity mode dynamics can also be adiabatically eliminated.
In particular, if $\kappa\gg |\Omega (t)|$
(but still with $\nu_x\gg\kappa$), then the equation
for the motional mode reduces to
\begin{equation} \label{QLE}
\dot{\hat{b}}_x \simeq 
- [\Gamma (t)+i\nu_x] \hat{b}_x + 
e^{i\phi_{\rm A}}\sqrt{2\Gamma (t)} \, e^{-i\nu_x t}
\hat{a}_{\rm in}(t)\; ,
\end{equation}
where we define
\begin{equation}
\Gamma (t)=\Omega (t)^2/\kappa .
\end{equation}
The motional dynamics thus reduces to that of a simple damped harmonic
oscillator, with the (possibly) time-dependent damping rate 
$\Gamma (t)$. However, the quantum noise operator appearing in 
(\ref{QLE}) corresponds to the {\em light field} incident upon the
cavity, and hence one realizes a
simple coupling of the motional mode to {\em propagating
light fields} external to the cavity. 
More precisely, from the input-output theory of optical
cavities \cite{Walls94,Gardiner00}, 
it can be shown that the cavity output 
field is given, under the present circumstances, by
\begin{equation} \label{aout}
\hat{a}_{\rm out}(t) \simeq - \, \hat{a}_{\rm in}(t) -
\sqrt{2\Gamma (t)} \; \tilde{b}_x(t) \, .
\end{equation}
where 
$\tilde{b}_x={\rm e}^{i\nu_xt}\hat{b}_x$, and we have set 
$\phi_{\rm A}=0$ for simplicity.
So, given a vacuum input field to the cavity, the output light field
is {\em determined by the motional state of the atom} confined inside the
cavity. In this way, nonclassical motional states can be converted into
nonclassical light fields; for example, entanglement between the
$x$-dimension motional mode and, say, the $z$-dimension motional mode
can be converted into entanglement between the $z$-dimension motional 
mode and the propagating output light field. This light field may then 
be coupled to another atom-cavity system to generate distributed 
motional state entanglement.

\subsection{
Motional state transfer between distant locations
}

Following work by Cirac {\em et al}. \cite{Cirac97} on the
transmission of a qubit between two nodes of a quantum network, 
it is shown in \cite{Parkins99} that if the output field from one
of our atom-cavity configurations is incident on a second such
atom-cavity configuration, with the coupling between systems being
{\em unidirectional}, then with suitably tailored laser pulses 
${\cal E}_{\rm A1}(t)$ and ${\cal E}_{\rm A2}(t)$ applied at 
the two sites (amounting essentially to impedance matching, such that 
all of the light exiting the first cavity is absorbed by the second
cavity) one may realize the motional state transfer
\begin{equation}
|\phi\rangle_x^{(1)} \otimes |0\rangle_x^{(2)} \rightarrow
|0\rangle_x^{(1)} \otimes |\phi\rangle_x^{(2)} \, ,
\end{equation}
where $|\phi\rangle_x$ is an {\em arbitrary} quantum state
describing the motion along the $x$-axis. Note that the cavity
fields begin and end the transfer in the vacuum state.

The scheme is able to operate in the regime
where $\Omega (t)$ and $\kappa$ are comparable (in which case the
transfer rate is essentially determined by $\kappa$), 
but again the analysis simplifies in the case where the cavity modes
can be adiabatically eliminated as above. 
In this case, a master equation for the reduced density matrix 
describing the motional states of the two atoms, $\tilde{\rho}_x$, 
can be derived (in the rotating frame) as
\cite{Parkins99}
\begin{eqnarray} \label{MEcascade}
\dot{\tilde{\rho}}_x &=& \Gamma_1(t)
\left( 2\tilde{b}_{1x}\tilde{\rho}_x\tilde{b}_{1x}^\dagger
- \tilde{b}_{1x}^\dagger\tilde{b}_{1x}\tilde{\rho}_x 
- \tilde{\rho}_x\tilde{b}_{1x}^\dagger\tilde{b}_{1x} \right) \nonumber
\\
&& + \Gamma_2(t) \left( 2\tilde{b}_{2x}\tilde{\rho}_x\tilde{b}_{2x}^\dagger
- \tilde{b}_{2x}^\dagger\tilde{b}_{2x}\tilde{\rho}_x 
- \tilde{\rho}_x\tilde{b}_{2x}^\dagger\tilde{b}_{2x} \right) \nonumber
\\
&& + 2\sqrt{\Gamma_1(t)\Gamma_2(t)} 
\left\{ [\tilde{b}_{2x}^\dagger ,\tilde{b}_{1x}\tilde{\rho}_x ]
e^{-i(\phi_{\rm 1A}-\phi_{\rm 2A})} \right. \nonumber
\\
&& \;\;\;\;\;\;\;\;\;\; \left.
+ [\tilde{\rho}_x\tilde{b}_{1x}^\dagger ,\tilde{b}_{2x} ] 
e^{i(\phi_{\rm A1}-\phi_{\rm A2})} \right\} \, .
\end{eqnarray}

Example pulse shapes for this regime, 
specified through the effective coupling rates 
of the motional modes to the external light fields,
$\Gamma_1(t)$ and $\Gamma_2(t)$, are 
(taking $\phi_{\rm A1}=\phi_{\rm A2}$) \cite{Parkins99}
\begin{equation}
\Gamma_1(t) = \Gamma \, \frac{{\rm e}^{\Gamma t}}
{{\rm e}^{\Gamma t}+{\rm e}^{-\Gamma t}} \, , \;\;\;\;
\Gamma_2(t) = \Gamma_1(-t) \, ,
\end{equation}
assuming the transfer starts at $t=-\infty$ and concludes at
$t=+\infty$, with $\Gamma$ a constant.
Armed with this capability, we are able to distribute quantum
states of a material oscillator, and generate entanglement,
between macroscopically-separated locations.

\subsection{
Numerical analysis
}

\subsubsection{
Damped harmonic oscillator model
}

The description of the motional mode dynamics in terms of a
linearly-damped harmonic oscillator coupled to propagating light
fields, Eq.(\ref{QLE}), represents a tremendous simplification
of the model and offers a very direct and transparent scheme 
for state transfer between motion and light. 
It is important then to gauge the validity of this simplification;
here we present some results from a numerical analysis of the
model starting from the Hamiltonian
\begin{eqnarray} \label{Hacsim}
\hat{H}_{\rm ac} && = 
\hbar\nu_x \left( \hat{b}_x^\dagger\hat{b}_x+1/2\right) + 
\hbar\delta_{\rm cA} \hat{a}^\dagger\hat{a} \nonumber
\\
&& \;\;\;  -\, \frac{\hbar {\cal E}_{\rm A}^2}
{\Delta_{\rm 0A}} - \frac{\hbar g_0^2}{\Delta_{\rm 0A}} \, 
\sin^2(k\hat{x}) \hat{a}^\dagger\hat{a}
\nonumber
\\
&& \;\;\; -\, \frac{\hbar g_0{\cal E}_{\rm A}}{\Delta_{\rm 0A}}\, 
\sin (k\hat{x}) \left(
e^{-i\phi_{\rm A}}\hat{a}^\dagger
+ e^{i\phi_{\rm A}}\hat{a} \right) ,
\end{eqnarray}
and master equation
\begin{eqnarray} \label{ME}
\dot{\hat{\rho}} = -\frac{i}{\hbar} [\hat{H}_{\rm ac},\hat{\rho}]
 + \kappa \left( 2\hat{a}\hat{\rho}\hat{a}^\dagger
-\hat{a}^\dagger\hat{a}\hat{\rho} - \hat{\rho}\hat{a}^\dagger\hat{a}
\right) ,
\end{eqnarray}
where the last term in (\ref{ME}) describes cavity damping.
The Hamiltonian (\ref{Hacsim}) is the one-dimensional form of
(\ref{Hac}) and neglects the position dependence of 
${\cal E}_{\rm A}$ (based on earlier arguments; see Appendix B).
These equations make no assumption about the Lamb-Dicke parameter in 
the $x$ direction and retain the dynamics of the cavity mode.

We solve the master equation (\ref{ME}) numerically using truncated
number state bases for the (harmonic oscillator) cavity and motional 
modes. The particular example we have concentrated on is the decay
of an initial coherent state of the motional mode, since this allows
a simple analytical solution in the ideal case where the motional mode
dynamics is exactly described by (\ref{QLE}). In particular, for an
initial coherent amplitude $\alpha$, the state evolves as
$|\psi (t)\rangle_x =e^{-i\nu_xt/2}|\alpha e^{-(i\nu_x+\Gamma )t}
\rangle_x$ in the ideal case.

A natural quantity to consider then in comparing the simulated evolution
with the ideal behavior is the fidelity
\begin{equation}
f(t) = {}_x\langle\alpha e^{-(i\nu_x+\Gamma )t}|\hat{\rho}_x|
\alpha e^{-(i\nu_x+\Gamma )t}\rangle_x ,
\end{equation}
where $\hat{\rho}_x={\rm Tr}_{\rm cav}\{\hat{\rho}\}$
is the reduced density operator for the motional mode and
\begin{equation}
\Gamma = \frac{1}{\kappa} \left( 
\frac{\eta_xg_o{\cal E}_{\rm A}}{\Delta_{\rm 0A}} \right)^2 .
\end{equation}

\paragraph{Results}

Results for two values of the Lamb-Dicke parameter,
$\eta_x=0.1$ and $\eta_x=0.15$, are shown in Fig.~4, where we plot
the magnitudes of the motional mode and cavity mode amplitudes as a
function of time, with the coherent state amplitude chosen to be
$\alpha =\sqrt{10}$.
In dimensionless units we choose $\kappa =1$,
$\nu_x=\delta_{\rm cA} =10$, $g_0^2/\Delta_{\rm 0A} =0.2$, 
and $\eta_xg_0{\cal E}_{\rm A}/\Delta_{\rm 0A} =0.1$ ($\ll\kappa$), 
corresponding to $\Gamma =0.01$.
In Fig.~4(a), $|\langle\hat{b}_x(t)\rangle |$ is seen to follow very 
closely the ideal behavior of a decaying coherent state, while Figs.~4(b,c)
demonstrate that the cavity mode, after an initial transient period,
adiabatically follows the motional mode. There is a small but noticeable
improvement (with respect to the ideal behavior) with the decrease in
$\eta_x$ from $0.15$ to $0.1$. Further improvements occur with smaller
$(\eta_xg_0{\cal E}_{\rm A}/\Delta_{\rm 0A})/\kappa$ (condition of 
adiabaticity) and larger $\nu_x/\kappa$ (rotating-wave approximation with 
respect to the trap frequency), as one would expect.

The fidelity $f(t)$ is plotted in Fig.~5, from which it is clear that, for
$\eta =0.1$, the motional state remains close to the desired state at all
times (at large $t$, of course, the states all approach the ground state).
For $\eta_x=0.15$ the deviation is more significant and the mapping of the
motional state onto the light field is evidently degraded.

\paragraph{Lamb-Dicke approximation}

Let us return briefly to some of the approximations made in deriving the
ideal model. If we consider the Hamiltonian (\ref{Hacsim}), the term 
proportional to $g_0^2/\Delta_{\rm 0A}$ 
is essentially negligible under the present
circumstances due to the smallness of $\eta_x$ 
and of the intracavity photon
number ($\langle\hat{a}^\dagger\hat{a}\rangle <0.1$ at early times 
and decreases with time). Consider now the expansion of $\sin (k\hat{x})$. 
To third order in $\eta_x$
\begin{eqnarray}
\sin (k\hat{x}) &=& \eta_x \left( \hat{b}_x + \hat{b}_x^\dagger \right) -
\frac{\eta_x^3}{3!} \left( \hat{b}_x + \hat{b}_x^\dagger \right)^3
\nonumber
\\
&=& \eta_x \left( 1 - \frac{\eta_x^2}{2} - 
\frac{\eta_x^2}{2}\hat{b}_x^\dagger
\hat{b}_x - \frac{\eta_x^2}{6}\hat{b}_x^2 \right) \hat{b}_x + {\rm h.c.} .
\end{eqnarray}
The reduction of the atom-cavity dynamics to a coupling of the form 
$\hat{a}^\dagger\hat{b}_x+{\rm h.c.}$ 
requires that the terms in the brackets
proportional to $\eta_x^2$ have negligible effect (compared to 1). The 
combination of small $\eta_x$ and rapid rotation means that the terms 
$\eta_x^2\hat{b}_x^2/6$ and $\eta_x^2\hat{b}_x^{\dagger 2}/6$ should be 
negligible, leaving a condition of the form
\begin{equation}
\frac{1}{2}\eta_x^2 (1+\bar{n}_x) \ll 1,
\end{equation}
where $\bar{n}_x\equiv \langle\hat{b}_x^\dagger\hat{b}_x\rangle$. 
This condition is reasonably well satisfied for the numerical examples 
considered, but a stronger condition would take into account the width of 
the number state distribution, i.e., the fact that vibrational population 
can reside in appreciable amounts in number states $|n\rangle_x$ such that 
$n>\bar{n}_x$. If $\sigma_{\bar{n}_x}^2$ is the variance of the 
number state distribution, then such a condition might take the form 
\begin{equation}
\frac{1}{2}\eta_x^2 (1+\bar{n}_x+a\sigma_{\bar{n}_x}) \ll 1,
\end{equation}
where $a\sim 2-3$ (i.e., several standard deviations from the mean).
For the coherent state considered above, 
$\sigma_{\bar{n}_x}=\sqrt{\bar{n}_x}
=\sqrt{10}$, leading to a condition (taking $a=3$)
\begin{equation}
\frac{1}{2}\eta_x^2 (1+\bar{n}_x+3\sigma_{\bar{n}_x}) 
\simeq 10\eta_x^2 \ll 1.
\end{equation}
With $\eta_x=0.15$, $10\eta_x^2=0.225$, and so one might 
expect that the Lamb-Dicke approximation starts to break down.

In the case of a two-mode squeezed state of the motion, 
the number state distribution
of the individual modes is akin to that of a thermal mode, for which 
$\sigma_{\bar{n}_x}=(\bar{n}_x^2+\bar{n}_x)^{1/2}\simeq\bar{n}_x+1/2$ for
$\bar{n}_x>1$. With $a=3$, we then require that 
\begin{equation}
\frac{1}{2}\eta_x^2 (4\bar{n}_x+5/2) \ll 1,
\end{equation}
which, for $\eta_x=0.1$, reduces to $\bar{n}_x=\sinh^2(r)\ll 50$.
This condition is reasonably well satisfied for values of $r$ up to 
$\sim 1.6$; beyond this, a smaller value of $\eta_x$ would be desirable.

\subsubsection{
Motional state transfer
}

The next problem we wish to examine in more detail is that of transferring 
a motional quantum state from one atom to another at a distant site using 
the procedure outlined in Section III.B and depicted schematically in 
Fig.~6. The coupling laser fields, and 
hence the equations of motion, now have an explicit time dependence.
This, combined with the increased dimensionality of the problem (now that
we have two atom-cavity systems to simulate), leads us to employ the
technique of Monte-Carlo wave function simulation (see, for example,
\cite{Gardiner00}).

In fact, the Monte-Carlo wave function technique is not only convenient, 
but also highly appropriate for modeling and analyzing the state transfer 
procedure as shown in Fig.~6. In this approach, the evolution of our 
quantum system is simulated by propagating a wave function 
$|\Psi (t)\rangle$ according to the Schr\"odinger equation 
\begin{equation} \label{SchrEqn}
\frac{\partial}{\partial t}
|\Psi (t)\rangle = -\frac{i}{\hbar}\hat{H}_{\rm eff}(t) |\Psi (t)\rangle ,
\end{equation}
where $\hat{H}_{\rm eff}(t)$ is a non-Hermitian effective Hamiltonian.
For the state-transfer configuration we are considering 
here, our effective Hamiltonian takes the form
\begin{eqnarray} \label{Heff}
\hat{H}_{\rm eff}(t) &=& \hat{H}_{\rm ac}^{(1)}(t) + 
\hat{H}_{\rm ac}^{(2)}(t) \nonumber
\\
&& -\, i\hbar\kappa_1\hat{a}_1^\dagger\hat{a}_1 - 
i\hbar\kappa_2\hat{a}_2^\dagger\hat{a}_2 - 2i\hbar\sqrt{\kappa_1\kappa_2}\,
\hat{a}_2^\dagger\hat{a}_1 ,
\end{eqnarray}
with
\begin{eqnarray} \label{Hacj}
\hat{H}_{\rm ac}^{(j)} (t) && =
\hbar\nu_{xj} \hat{b}_{xj}^\dagger\hat{b}_{xj} + 
\hbar\delta_{\rm cA}^{(j)}
\hat{a}_j^\dagger\hat{a}_j \nonumber
\\
&& -\, \frac{\hbar g_{0j}^2}{\Delta_{\rm 0A}} \, \sin^2(k\hat{x}_j) 
\hat{a}_j^\dagger\hat{a}_j
\nonumber
\\
&& -\, \frac{\hbar g_{0j}{\cal E}_{{\rm A}j}(t)}{\Delta_{\rm 0A}}\, 
\sin (k\hat{x}_j) \left(
e^{-i\phi_{{\rm A}j}}\hat{a}_j^\dagger
+ e^{i\phi_{{\rm A}j}}\hat{a}_j \right) .
\end{eqnarray}
[For simplicity, we now omit constant energy shifts from the 
Hamiltonians $\hat{H}_{\rm ac}^{(j)}(t)$, and the frequencies of the two
coupling lasers are assumed to be the same ($\omega_{\rm A}$).] 
The second line in
(\ref{Heff}) follows from the cascaded-systems formalism in which one 
assumes a unidirectional coupling between the cavities
\cite{Gardiner00,Gardiner93,Carmichael93}.

The propagation described by (\ref{SchrEqn})
is interrupted at random times $\{ t_r\}$ 
by wave function collapses, or quantum jumps, 
\begin{equation}
|\Psi (t_r+dt)\rangle = \frac{\hat{C}|\Psi (t_r)\rangle}
{\langle\Psi (t_r)|\hat{C}^\dagger\hat{C}|\Psi (t_r)\rangle^{1/2}} 
\end{equation}
(which include renormalization of the wave function), 
where $\hat{C}$ is an appropriate
jump operator and the probability density for a jump to occur during the
time interval from $t$ to $t+dt$ is 
$\langle\Psi (t)|\hat{C}^\dagger\hat{C}|\Psi (t)\rangle dt$. 
For our situation the jump operator is
\begin{equation}
\hat{C} = \sqrt{\kappa_1}\,\hat{a}_1 + \sqrt{\kappa_2}\,\hat{a}_2 ,
\end{equation}
which can be identified with the destructive detection of a photon by 
the photodetector monitoring the output channel from the second cavity.

For ideal quantum transmission we require that a quantum jump 
(that is, a photon detection) never occurs, i.e., 
$\hat{C}|\Psi (t)\rangle =0$ for all times $t$ (which also means that the 
norm of the wave function remains equal to 1 at all times).
In other words, all of the 
quantum information is transferred from atom 1 to atom 2 and none is lost
to light fields propagating away from the system. 
The laser pulse profiles derived from the simplified master equation model 
(\ref{MEcascade}) should approximately satisfy this condition and facilitate
high-fidelity state transfer, provided the various parameters of the system 
are chosen appropriately.
Here we want to assess more carefully the performance of these profiles for
a more comprehensive model of the atom-cavity dynamics, as described by the 
Hamiltonians in (\ref{Hacj}). 

We assume an initial state of the form 
\begin{equation}
|\Psi (t=-\infty )\rangle = |\phi\rangle_{x}^{(1)} \otimes 
|0\rangle_{\rm cav}^{(1)} \otimes |0\rangle_{\rm cav}^{(2)} \otimes
|0\rangle_{x}^{(2)} ,
\end{equation}
where $|\phi\rangle_{x}^{(1)}$ is the particular motional quantum state to
be transmitted. The target state of the transmission is then
\begin{equation}
|\Psi (t=+\infty )\rangle_{\rm target} = |0\rangle_{x}^{(1)} \otimes 
|0\rangle_{\rm cav}^{(1)} \otimes |0\rangle_{\rm cav}^{(2)} \otimes
|\phi\rangle_{x}^{(2)} .
\end{equation}
For simplicity, we will also assume identical atom-cavity systems, i.e.,
\begin{eqnarray}
&& \nu_{x1} = \nu_{x2} \equiv \nu_x , \;\;\;
\eta_{x1} = \eta_{x2} \equiv \eta_x , \nonumber
\\
&& g_{01} = g_{02} \equiv g_0 , \;\;\; \kappa_1 = \kappa_2 \equiv \kappa ,
\;\;\; \delta_{\rm cA}^{(1)} = \delta_{\rm cA}^{(2)} \equiv \delta_{\rm cA} .
\end{eqnarray}

\vspace{3mm}

\paragraph{Laser pulse profiles}

The time dependence of the laser field ${\cal E}_{\rm A1}(t)$ is chosen 
to satisfy
\begin{equation}
\frac{1}{\kappa} \left[ \frac{\eta_xg_0{\cal E}_{\rm A1}(t)}{\Delta_{\rm 0A}}
\right]^2 \equiv \Gamma_1(t) = \Gamma \frac{e^{\Gamma t}}{e^{\Gamma t}+
e^{-\Gamma t}} ,
\end{equation}
with 
\begin{equation}
\Gamma = \frac{1}{\kappa} 
\left[ \frac{\eta_xg_0{\cal E}_{\rm A1}^{\rm max}}{\Delta_{\rm 0A}}
\right]^2 ,
\end{equation}
while ${\cal E}_{\rm A2}(t)$ is chosen such that
$\Gamma_2(t)=\Gamma_1(-t)$. Note again that these example 
forms for the temporal profiles 
of the laser fields are derived from a theoretical analysis in which 
the dynamics of the cavity modes are adiabatically eliminated, and hence,
depending on the choices of parameters, they may not be optimal choices.
We will return to this point when we present the numerical results.

\paragraph{Truncation of trigonometric functions}

To help speed up our computations we have in fact performed most of
our simulations using expansions of the operators $\sin^2(k\hat{x}_j)$ and
$\sin (k\hat{x}_j)$ that are truncated 
to third order in the Lamb-Dicke parameter $\eta_x$.
For the range of parameters considered here, we find little difference 
between results obtained using the full trigonometric forms and those
using the truncated versions.
With the truncated versions it is also straightforward to move to a
rotating frame so as to remove the systematic (fast) evolution given 
by the first line in (\ref{Hacj}).
This does of course result in rapidly rotating factors multiplying some of
the remaining terms of the Hamiltonians [in particular, factors of the forms
$e^{\pm i(\nu_x+\delta_{\rm cA})t}$ or $e^{\pm i(3\nu_x\pm\delta_{\rm cA})t}$], 
and we retain these terms in the simulations.

\paragraph{Fidelity of the transmission}

In the context of our wave function simulations, the ideal situation
corresponds to the case where, following the laser pulse sequence, the
norm of the wave function is still equal to 1, and hence no jumps have
occurred. However, 
due to non-ideal operating conditions (for example, rapidly rotating
``off-resonant'' terms making a finite contribution to the dynamics)
the norm of the wave function does decay and hence there is a finite 
possibility of a photon detection and consequent loss of information.
Now, in a typical application of the Monte-Carlo wave function approach
one averages over many trajectories to obtain a density operator 
for the system. From this density operator one could, in the present
context, compute the average fidelity for the quantum state transmission.

However, the decay of the norm in a single trajectory already provides 
us with a good indicator of the performance of the transmission.
If, for example, during the pulse sequence and state transfer the 
norm decays to a value of 0.9 (without any jumps occurring), 
then we can make the general statement that in 
90\% of our attempts the state will be transferred successfully.  
Note that, in the ``no-jump'' case, on renormalizing the simulated wave 
function $|\Psi (t_{\rm f})\rangle_{\rm sim}$ (where $t_{\rm f}$ is the 
finishing time for the simulation) we find that the fidelity of
the transmission, defined to be
\begin{equation}
F = \left|\, {}_{\rm sim}\langle\Psi(t_{\rm f})|\Psi (+\infty)
\rangle_{\rm target} \,\right|^2 ,
\end{equation}
is very close to 1 (i.e., $\geq 0.99$)
for all of the numerical examples we consider below. 
The value of 0.9 then also essentially
sets a {\em lower bound} on the {\em average fidelity} 
of the state transmission. In the 10\% of cases 
where one or more quantum jumps do occur the final state
$|\Psi (t_{\rm f})\rangle_{\rm sim}$ may, depending on the precise nature
of the state being transferred, still have a finite 
(and even substantial) overlap with the
target state, and so the average fidelity will actually be larger than 0.9.

From a practical point of view, it is also worth noting that it would 
in principle be possible to post-select high-fidelity state transmissions 
by monitoring the output from the second cavity. If a photon is {\em not} 
detected in this output then the transmission is known to have been 
successful (assuming perfect detection efficiency). If a photon is detected 
then the system can be reset and the transfer attempted again (or some 
form of error correction procedure could be applied).

\paragraph{Results}

As the state to be transferred, we have concentrated on the following 
examples:
(i) a truncated phase state,
\begin{equation}
|\phi\rangle_x^{(1)} = \frac{1}{\sqrt{N+1}} 
\sum_{n=0}^N |n\rangle_x^{(1)} ,
\end{equation}
i.e., a uniformly-weighted coherent superposition of the first $N+1$ 
Fock states,
(ii) a pure Fock state 
\begin{equation}
|\phi\rangle_x^{(1)} = |N\rangle_x^{(1)} ,
\end{equation}
and (iii) a Schr\"odinger Cat state
\begin{equation}
|\phi\rangle_x^{(1)} = \frac{1}{{\cal N}_+} \, \left( 
|\alpha\rangle_x^{(1)} + |-\alpha\rangle_x^{(1)} \right) ,
\end{equation}
with $|\alpha\rangle_x^{(1)}$ a coherent state
and ${\cal N}_+=[2(1+e^{-2|\alpha |^2})]^{1/2}$.

In dimensionless units, we again set $\kappa =1$ and choose 
$g_0^2/\Delta_{\rm 0A}=0.2$ and 
$(g_0{\cal E}_{\rm A}^{\rm max})/\Delta_{\rm 0A}=1$. For $\eta_x=0.1$
this gives 
$(\eta_xg_0{\cal E}_{\rm A}^{\rm max})/\Delta_{\rm 0A}=0.1\kappa$
and so the adiabatic approximation
[$(\eta_xg_0{\cal E}_{\rm A}^{\rm max})/\Delta_{\rm 0A}\ll\kappa$] 
used in deriving the laser pulse shapes should be reasonably good. 
Indeed, we find that the improvement in the fidelity of the transfer one 
obtains when $\eta_x$ is reduced from $0.1$ to $0.0707$ (see tables below)
results primarily from an improvement in the Lamb-Dicke approximation 
rather than in the adiabatic approximation.

\vspace{2mm}
\noindent
(i) Truncated phase state:
Some results illustrating the performance of the transfer for
this state are presented in Tables 
\ref{table2} ($N=10$) and \ref{table3} ($N=20$). 
We have considered various combinations 
of the trapping frequency $\nu_x$ and the Lamb-Dicke parameter $\eta_x$.
The third column gives the norm of the wave function at the conclusion of
the transfer operation (with quantum jumps ``turned off''), 
which, as discussed earlier, effectively
amounts to a lower bound on the average transfer fidelity, $F_{\rm ave}$.

As one can see, a large trapping frequency is very important for obtaining
a high average transfer fidelity. In particular, a value 
$\nu_x/\kappa\geq 10$ is necessary if one wishes to obtain a success rate
exceeding $\sim 80$\% for the case $N=20$.
As the dimensionality of the motional state being transferred increases
(i.e., higher Fock states are populated),
the size of $\eta_x$ also becomes more critical.
This can be seen by comparing the results in Tables \ref{table2} and 
\ref{table3}. Smaller values of $\eta_x$ are clearly required (for a
given value of $\nu_x$) in order for the state
$1/\sqrt{21}\sum_{n=0}^{20}|n\rangle_x$ to be transferred with a fidelity
comparable to that for the state $1/\sqrt{21}\sum_{n=0}^{10}|n\rangle_x$.
The relative improvement in performance with a decrease is $\eta_x$ is also
more pronounced for the state of higher dimensionality.

With the inclusion of quantum jumps some interesting and complicated 
behavior is observed in the individual trajectories. For the 
phase state considered the overlap of the transmitted state with the
ideal state can be substantial ($\sim 0.40-0.75$), but the precise
nature of the transmitted state depends crucially on the time at which 
each jump occurs.

\vspace{2mm}
\noindent
(ii) Fock state:
Results for transmission of the Fock state $|n=10\rangle_x$ are given in 
Table \ref{table4}. The performance is in fact comparable to that for the 
phase state $1/\sqrt{21}\sum_{n=0}^{20}|n\rangle_x$, which actually has 
the same mean phonon number of 10. 

The effect of quantum jumps on the transfer is,
as one would expect, more severe than in the case of the phase state; 
overlap with the state $|n=10\rangle_x$ disappears completely with a 
single photon detection. However, after the jump the resulting final 
state is not necessarily  what one might naively expect (i.e., the state
$|n=9\rangle_x$). It is possible,  as a result of the finite effects of
``non-resonant'' terms in the dynamics, for the transferred state to take 
the form of a superposition of the states
$|n=9\rangle_x$ and $|n=11\rangle_x$.

\vspace{2mm}
\noindent
(iii) Schr\"odinger Cat state:
Finally, we consider the transfer of a mesoscopic superposition of coherent
states of the form ${\cal N}_+^{-1}(|\alpha\rangle_x+|-\alpha\rangle_x)$,
with $\alpha =\sqrt{10}$. This state also has a mean excitation (phonon) 
number equal to 10 and the results shown in Table \ref{table5} are very 
similar to those of the previous two examples. 

The effect of quantum jumps is particularly interesting in this case, as
we find that, with a single photon detection, the transferred state is in 
general very close to the state 
${\cal N}_-^{-1}(|\alpha\rangle_x-|-\alpha\rangle_x)$ (i.e., an ``odd'' 
Schr\"odinger Cat state), 
suggesting that an operation on the motional state amounting
to an application of the annihilation operator $\hat{b}_{x2}$ would largely
restore the original state. 
Note that the even and odd Schr\"odinger Cat states 
${\cal N}_+^{-1}(|\alpha\rangle_x+|-\alpha\rangle_x)$ and
${\cal N}_-^{-1}(|\alpha\rangle_x-|-\alpha\rangle_x)$
have been proposed, in the context of quantum computation, 
as logical qubit encodings for the correction of
bit-flip errors caused by amplitude damping (as occurs in our system) 
\cite{Cochrane99}. 
They have also been proposed for use in a secure quantum key distribution 
protocol \cite{Nambu00}.

\paragraph{Atomic spontaneous emission}

While our model and simulations have included cavity damping, we have until
now ignored any effects associated with atomic spontaneous emission, which
arises from the small but finite probability for the atom to be in its
excited internal state $|e\rangle$. In Appendix A (Part 2) an approximate 
expression is derived for the rate at which atomic spontaneous emission is 
expected to influence the motional state of the atom. In order for this rate
to be much slower than the characteristic state transfer rate $\Gamma$ of
the present configuration, one requires that the condition
\begin{equation}
\frac{10g_0^2}{\kappa\gamma} \gg 1
\end{equation}
be satisfied, where $\gamma$ is the atomic spontaneous emission rate. Hence,
one desires the regime of strong-coupling cavity QED, for which
$g_0^2/(\kappa\gamma )\gtrsim 1$.

\paragraph{Experimental prospects}

Let us focus again on the case of trapped ${}^9{\rm Be}^+$ ions, although 
other ionic or atomic species could evidently be considered (and might
perhaps be preferable from the point of view of cavity QED experiments). 
The ${}^2{\rm S}_{1/2}\leftrightarrow {}^2{\rm P}_{1/2}$ transition 
wavelength is $\lambda =313$~nm and the transition linewidth is
$\gamma /(2\pi )=19.4$~MHz. 
If we assume, for example, 
that the mirrors forming the cavity have radii of curvature
equal to 5~cm and are separated by a distance $l=1\,(2)$~mm, then 
$g_0/(2\pi )=5.3\,(3.1)$~MHz. For a cavity finesse of 75,000 one obtains
$\kappa /(2\pi )=1.0\,(0.5)$~MHz, and so $10g_0^2/(\kappa\gamma )=14\,(10)$, 
while a trap frequency of $\nu_x/(2\pi )=22$~MHz
(corresponding to a Lamb-Dicke parameter $\eta_x=0.1$) 
gives $\nu_x/\kappa =22\,(44)$.
With these parameters one would anticipate a state transfer rate 
$\Gamma /(2\pi )\sim 10-20$~kHz. Note that the timescales for motional
decoherence and heating observed in recent trapped ion experiments are of
the order of milliseconds or longer \cite{Wineland98a,Roos99,Turchette00}.

There are of course many other possible combinations of parameters which
should satisfy (at least approximately) the essential requirements of the
scheme. Much larger cavity finesses may be possible, but challenging, at this
wavelength, allowing smaller mirror separation and larger dipole coupling
strength $g_0$ (which increases as the cavity mode volume decreases).
However, it may be difficult in practice to bring mirrors very close together
about an ion trap (for example, because of the technical problem of charge 
build-up on the mirrors).
For this reason, it is perhaps advantageous to consider neutral atom
experiments (in particular, alkali atom experiments), where, in fact, 
optical dipole traps (see, for example, \cite{Ye99})
or microscopic magnetic traps (see, for example, \cite{Fortagh00})
should allow confinement of atoms in the Lamb-Dicke regime with trapping 
frequencies also in the MHz range. Furthermore, cavity QED experiments 
with alkali atoms are already highly developed (for recent experiments,
see \cite{Ye99,Hood00,Pinkse00}) and spectacular cavity finesses can be 
achieved at the relevant (longer) wavelengths \cite{Rempe92}.

\subsection{
Multiple cavity-confined atoms 
}

It is interesting to consider the possibility of having more than one atom
coupled to the cavity field at a time. The kind of situation one might 
imagine is like that realized in the recent experiment of Ye {\em et al}.
\cite{Ye99},
in which (single) atoms were trapped inside a microscopic optical cavity
using a far-off-resonance dipole-force trap (FORT). The FORT was actually
produced by excitation of an ``auxiliary'' longitudinal cavity mode, and 
so had a standing-wave structure colinear with (but of slightly different
periodicity to) the cavity QED mode of interest. Hence, given sufficiently
strong confinement, the FORT can be regarded as a chain of individual
cavity-confined microtraps, each of which could in principle be occupied
by a different atom.

With, for simplicity,
a single coupling field ${\cal E}_{\rm A}$ incident on the atoms from the
side of the cavity, a Hamiltonian for the system can be written as
(assuming a single microtrap frequency $\nu_x$, and tight confinement along 
transverse directions as well)
\begin{eqnarray} \label{HacNatom}
\hat{H}_{\rm ac} &&= \sum_j\hbar\nu_x\left( \hat{b}_{xj}^\dagger\hat{b}_{xj}
+ 1/2 \right) + \hbar\delta_{\rm cA} \hat{a}^\dagger\hat{a} \nonumber
\\
&& - \sum_j \frac{\hbar g_0^2}{\Delta_{\rm 0A}} \sin^2(k\hat{x}_j+\theta_j) 
\hat{a}^\dagger \hat{a} - 
\sum_j \frac{\hbar {\cal E}_{\rm A}^2}{\Delta_{\rm 0A}} 
\nonumber
\\
&& - \sum_j \frac{\hbar g_0{\cal E}_{\rm A}}{\Delta_{\rm 0A}} 
\sin (k\hat{x}_j+\theta_j)
\left( e^{-i\phi_{\rm A}}\hat{a}^\dagger + e^{i\phi_{\rm A}}\hat{a} \right) ,
\end{eqnarray}
where $\theta_j$ defines the position of the center of the $j$-th microtrap
relative to a node of the cavity QED field.

In the Lamb-Dicke limit we can write
\begin{eqnarray}
\sin (k\hat{x}_j+\theta_j) &=& \sin (k\hat{x}_j)\cos (\theta_j) + 
\cos (k\hat{x}_j)\sin (\theta_j) \nonumber
\\
&\simeq & \eta_x \left( \hat{b}_{xj} + \hat{b}_{xj}^\dagger \right) \cos 
(\theta_j) + \sin (\theta_j) .
\end{eqnarray}
Assuming that $g_0^2/\Delta_{\rm 0A}\ll\delta =\nu_x$ and that 
$g_0\langle\hat{a}^\dagger\hat{a}\rangle^{1/2}\ll {\cal E}_{\rm A}$, 
we can neglect
the term in (\ref{HacNatom}) proportional to $g_0^2/\Delta_{\rm 0A}$, 
while in the last term we can neglect the contribution proportional to 
$\sin (\theta_j)(e^{-i\phi_{\rm A}}\hat{a}^\dagger 
+e^{i\phi_{\rm A}}\hat{a})$
as this term is rapidly rotating (at frequencies $\pm\delta_{\rm cA}$, 
with 
$\delta_{\rm cA}=\nu_x\gg\kappa ,g_0{\cal E}_{\rm A}/\Delta_{\rm 0A}$) 
compared to terms of the
form $\hat{a}^\dagger\hat{b}_{xj}$ and $\hat{b}_{xj}^\dagger\hat{a}$.

Consequently, we can describe the dynamics of the system through equations 
of the form
\begin{eqnarray}
\dot{\hat{a}} &=& -(\kappa +i\delta_{\rm cA})\hat{a} \nonumber
\\
&& - ie^{-i\phi_{\rm A}}\sum_j
\Omega_j\hat{b}_{xj} - \sqrt{2\kappa} e^{-i\nu_xt}\hat{a}_{\rm in}(t) ,
\\
\dot{\hat{b}}_{xj} &=& -i\nu_x\hat{b}_{xj} - ie^{i\phi_{\rm A}}
\Omega_j\hat{a} ,
\end{eqnarray}
with 
\begin{equation}
\Omega_j = -\frac{\eta_xg_0{\cal E}_{\rm A}}{\Delta_{\rm 0A}} 
\cos (\theta_j) .
\end{equation}
Adiabatic elimination of the cavity mode as before then leads to
\begin{eqnarray}
\dot{\hat{b}}_{xj} &=& -(\Gamma_j+i\nu_x)\hat{b}_{xj} - 
\sum_{k\neq j} \frac{\Omega_j\Omega_k}{\kappa} \hat{b}_{xk} \nonumber
\\
&& + e^{i\phi_{\rm A}} \sqrt{2\Gamma_j} e^{-i\nu_xt} \hat{a}_{\rm in}(t) ,
\end{eqnarray}
where $\Gamma_j=\Omega_j^2/\kappa$. So, through the laser and cavity fields
one realizes a coupling between the motional modes of different atoms.

Now, defining the ``collective'' mode operator
\begin{equation}
\hat{B}_x = \frac{1}{\sqrt{N_{\rm eff}}} \sum_k \cos (\theta_k) 
\hat{b}_{xk} ,
\end{equation}
with $N_{\rm eff}=\sum_j\cos^2(\theta_j)$ (such that 
$[\hat{B}_x,\hat{B}_x^\dagger ]=1$), one finds
\begin{equation} \label{Bmodedamp}
\dot{\hat{B}_x} = -(N_{\rm eff}\Gamma +i\nu_x) \hat{B}_x - 
\sqrt{2N_{\rm eff}\Gamma} e^{-i\nu_xt} \hat{a}_{\rm in}(t) ,
\end{equation}
with $\Gamma =(\eta_xg_0{\cal E}_{\rm A}/\Delta_{\rm 0A})^2/\kappa$.
This points to the possibility of using the collective mode for quantum 
state transfer and storage \cite{Bcool}. 
A potential advantage of such an approach is the
enhanced effective transfer rate $N_{\rm eff}\Gamma$, which (for
$N_{\rm eff}>1$) should enable the requirement of strong-coupling 
cavity QED to be relaxed somewhat.

The possibility also arises for the preparation of some very interesting
many-atom entangled states. For example, motional state transfer of 
a Schr\"odinger Cat state 
${\cal N}_+^{-1}(|\alpha\rangle_x+|-\alpha\rangle_x)$
to the collective mode, perhaps from a single-atom (cavity QED) source, 
would result in a many-atom state of the form (omitting the normalization)
\begin{equation}
|\alpha_1,\alpha_2,\alpha_3,\ldots\rangle_x +
|-\alpha_1,-\alpha_2,-\alpha_3,\ldots\rangle_x ,
\end{equation}
where
$\alpha_j=\alpha\cos (\theta_j)/\sqrt{N_{\rm eff}}$.

\subsection{
Coupling to collective trapped ion modes
}

Earlier, we discussed a scheme for coupling orthogonal single-ion and 
many-ion collective vibrational modes using a pair of laser fields. 
This scheme could also be utilized in the context of state transfer and 
storage, albeit somewhat indirectly, as depicted in Fig.~7.
In particular, state transfer between motion and light and from one
cavity location to another could be achieved using a single-ion 
interaction with the field mode (assuming that individual addressing of 
the ion by the coupling laser is possible), 
after which the ion-cavity coupling is switched off and the coupling 
between the single-ion mode and a particular collective motional mode
is switched on (for example, to transfer a quantum state from the 
single-ion mode to the collective mode). Storage of quantum information 
in certain collective vibrational modes could be advantageous due to the 
very slow heating rates  experienced by these modes \cite{King98}.

\section{
Entangling distant atoms
}

Having described and analysed basic techniques for preparing ``local''
entanglement of vibrational modes, and then for coupling vibrational 
modes to propagating light fields, we now want to outline some specific 
and interesting
possibilities for distribution of entanglement between distant locations.
The procedure is straightforward: one prepares an entangled state of the 
orthogonal vibrational modes in the $x$- and, say, $z$-directions, after 
which the $x$ mode is coupled to the cavity and thence to the external 
light field. With appropriate time-dependent coupling we have seen that 
the properties of the $x$ mode can be transferred with high fidelity to 
the corresponding mode of a distant cavity-confined atom (or atoms). 
That is, we can perform the transformation
\begin{eqnarray}
&& \left( \sum_{n,m} c_{nm} |n\rangle_x^{(1)}\otimes |m\rangle_z^{(1)} 
\right) \otimes |0\rangle_x^{(2)} \nonumber
\\
&& \;\;\;\; \rightarrow |0\rangle_x^{(1)} \otimes
\left( \sum_{n,m} c_{nm} |n\rangle_x^{(2)} \otimes |m\rangle_z^{(1)} 
\right) ,
\end{eqnarray}
where $|n\rangle_{x,z}^{(i)}$ is a vibrational Fock state of the atom 
at location $i$. Hence, the $x$ and $z$ modes of the atoms at locations 
2 and 1, respectively, become entangled.

Note that theoretical proposals exist for schemes that would in 
principle allow the preparation of {\em arbitrary} two-dimensional 
vibrational states of a  trapped atom 
\cite{Gardiner97,Drobny97,Drobny98,Kneer98}. It follows that, 
using the state transfer scheme, it would in principle be possible to
prepare an {\em arbitrary} entangled state of distantly-separated atoms. 
Below we highlight briefly just a few examples of particular interest, 
with emphasis on possibilities associated with the particular two-mode 
entangling schemes presented in Section II.

\subsection{
Examples
}

\subsubsection{
EPR state preparation between distant atoms
}

Here, we use the scheme of Section II.A.2 to prepare the state 
$|\psi_{\rm sq}(r)\rangle^{(1)}$ given in (\ref{psixz}), following which 
the state transfer operation outlined above produces the delocalized 
state 
\begin{equation}
\cosh^{-1}(r) \sum_{m=0}^\infty [-\tanh (r)]^m |m\rangle_x^{(2)} 
\otimes |m\rangle_z^{(1)} .
\end{equation}
The corresponding Wigner function is
\begin{eqnarray}
&& W\left(\tilde{x}^{(2)},\tilde{p}_x^{(2)};\tilde{z}^{(1)},
\tilde{p}_z^{(1)}\right)
\nonumber
\\
&& \; = \frac{4}{\pi^2} \exp \left\{ -\left[ ( \tilde{x}^{(2)}+
\tilde{z}^{(1)} )^2 + ( \tilde{p}_x^{(2)}-\tilde{p}_z^{(1)} )^2 \right]
e^{+2r} \right\} \nonumber
\\
&& \;\;\;\;\times \exp \left\{ -\left[ 
( \tilde{x}^{(2)}-\tilde{z}^{(1)}
)^2 + ( \tilde{p}_x^{(2)}+\tilde{p}_z^{(1)} )^2 \right]
e^{-2r} \right\} ,
\end{eqnarray}
where $\{\tilde{z}^{(1)},\tilde{p}_z^{(1)}\}$ and 
$\{\tilde{x}^{(2)},\tilde{p}_x^{(2)}\}$ are position and momentum 
variables for the atoms at locations 1 and 2, respectively.

So, one prepares an EPR state in position and momentum of a pair of 
(distantly) separated atoms. Apart from being of great historical 
significance, this state also constitutes the essential resource for 
the teleportation of continuous variables
\cite{Furusawa98,Vaidman94,Braunstein98}.
In particular, in the present context it offers the possibility of 
teleporting atomic center-of-mass wave functions between distant sites 
\cite{Parkins00b}.

\subsubsection{
Delocalized mesoscopic states
}

As mentioned earlier, the linear mixing operation described by (\ref{mixer})
can also generate entanglement given suitable initial states. Consider, for
example, the situation in which, say, the $z$ mode is initially prepared in 
its ground state while the $x$ mode is prepared in a Schr\"odinger Cat 
state, i.e.,
\begin{equation}
|\psi (0)\rangle = \frac{1}{{\cal N}_+} \left( |\alpha\rangle_x^{(1)}
+ |-\alpha\rangle_x^{(1)} \right) \otimes |0\rangle_z^{(1)} .
\end{equation}
Application of the linear mixing operation for a time $T=\pi /(4\chi )$
will transform this into the state
\begin{eqnarray}
|\psi (T)\rangle &=& \frac{1}{{\cal N}_+} \left( 
\left|\frac{\alpha}{\sqrt{2}}\right\rangle_x^{(1)}  \otimes
\left|\frac{-\alpha}{\sqrt{2}}\right\rangle_z^{(1)} 
\right. \nonumber 
\\
&& \;\;\;\;\;\;\;\;\;\; + \left.
\left|\frac{-\alpha}{\sqrt{2}}\right\rangle_x^{(1)} \otimes
\left|\frac{\alpha}{\sqrt{2}}\right\rangle_z^{(1)} 
\right) .
\end{eqnarray}
The state transfer procedure then generates the delocalized state
\begin{equation} \label{delocoh}
\frac{1}{{\cal N}_+} \left( 
\left|\frac{\alpha}{\sqrt{2}}\right\rangle_x^{(2)}  \otimes
\left|\frac{-\alpha}{\sqrt{2}}\right\rangle_z^{(1)} +
\left|\frac{-\alpha}{\sqrt{2}}\right\rangle_x^{(2)} \otimes
\left|\frac{\alpha}{\sqrt{2}}\right\rangle_z^{(1)} 
\right) .
\end{equation}
From what we have seen in our numerical analysis, the coherent state
amplitude $\alpha$ can be reasonably large
(while still allowing high fidelity operations and transfers), and so 
the entangled state (\ref{delocoh}) could be regarded 
as a delocalized {\em mesoscopic} state \cite{Parkins00c}. 

Schemes have also been proposed for preparing a state of the form 
\begin{equation} 
\frac{1}{\sqrt{2}} \left( |N\rangle_x^{(1)} \otimes |0\rangle_z^{(1)} +
|0\rangle_x^{(1)} \otimes |N\rangle_z^{(1)} \right) ,
\end{equation}
where $|N\rangle_x^{(1)}$ is a Fock state 
(see, for example, \cite{Wineland98b}), which would lead to a
delocalized state
\begin{equation} \label{delocN}
\frac{1}{\sqrt{2}} \left( |N\rangle_x^{(2)} \otimes |0\rangle_z^{(1)} +
|0\rangle_x^{(2)} \otimes |N\rangle_z^{(1)} \right) .
\end{equation}
Such states are of potential interest in the context of 
phase sensitivity in a two-mode interferometer, where they should allow
measurements at the Heisenberg uncertainty limit (see, for example,
\cite{Wineland98b}, and references therein).

\subsubsection{
Many-atom entangled states
}

Through the many-atom configurations discussed in Sections III.D and III.E
one can generalize the entanglement distribution procedure to collections 
of trapped atoms or ions located at the distant sites. For example, one
could imagine entangling {\em collective vibrational} modes of 
two strings of trapped ions at separate locations.
Such schemes complement proposals for entangling {\em collective internal} 
atomic states of separated free-space atomic ensembles using propagating 
light fields (see, for example, \cite{Polzik99,Lukin00,Kozhekin00,Duan00}).

\section{
Conclusion
}

In this work we have described and analysed schemes (i) for producing
motional state entanglement of an atom (or atoms) at one location and
(ii) for distributing this entanglement between atoms at distant
locations. The particular example we have focussed on for generating
(local) entanglement was chosen for its relative simplicity and for its
close relation to schemes already implemented in the laboratory. 
A further motivating factor which should also be emphasized is that
the operations that it allows are of direct relevance to quantum 
communication and computing with continuous variables, for which 
squeezers and linear mixers are basic elements (see, for example,
\cite{Furusawa98,Braunstein98,Lloyd99,Loock00,Braunstein00,Ralph00}). 

The cavity-QED-based 
motional state transfer scheme that enables the entanglement
to be distributed between distant sites has been introduced elsewhere
\cite{Parkins99}, but the present work extends significantly the numerical
analysis of this scheme. In particular, we have considered the transfer
of what may be regarded as mesoscopic motional states and from this
we have been able to gauge the validity of some of the more fundamental
assumptions implicit in the scheme, such as the Lamb-Dicke and
rotating-wave approximations.
Our calculations suggest that states of a substantial ``size'' can be
transferred with high fidelity for physically reasonable parameter values
and we have discussed some possible experimental scenarios,
encouraged by spectacular recent advances in atom and ion trapping 
technology and in cavity QED. 

The possibilities offered by the schemes for distributed entanglement
are many and varied, and we have discussed just a few examples in 
Section IV, along with some possible applications such as quantum
teleportation of atomic wavepackets. 
Of course, in addition to potential applications in quantum communication 
and computing, entangled and delocalized states of the form considered
here would offer some unique and fascinating opportunities for
fundamental tests of quantum mechanics versus local realism using 
{\em massive particles} \cite{Parkins00a,Parkins00c,Fry00}.

\acknowledgments
ASP thanks H.J. Kimble, H. Ritsch, I. Cirac, and D. Leibfried for 
discussions and comments
and gratefully acknowledges support from the Marsden Fund 
of the Royal Society of New Zealand. ASP also thanks the Quantum Optics
groups at the University of Innsbruck and the California Institute
of Technology for support and hospitality during visits when part 
of this work was carried out.

\appendix

\section{
Effects of atomic spontaneous emission
}

\subsection{
Motional state preparation
}

Clearly, a very important assumption is that the effects of 
atomic spontaneous emission can be neglected. 
In a master equation approach, atomic spontaneous emission with the
effects of recoil taken into account is modelled by a term of the
form (considering, for simplicity, motion only along the $x$ axis) 
\cite{Cirac92}
\begin{equation} \label{SpEme}
\{\dot{\hat{\rho}}\}_{\rm spon} = \frac{\gamma}{2} 
\left( 2\hat{\sigma}_-\tilde{\rho}\hat{\sigma}_+ - 
\hat{\sigma}_+\hat{\sigma}_-\hat{\rho} - 
\hat{\rho}\hat{\sigma}_+\hat{\sigma}_- \right) \, ,
\end{equation}
where
\begin{eqnarray}
\tilde{\rho} &=& \frac{1}{2} \int_{-1}^{+1} du \; W(u)
e^{iku\hat{x}} \hat{\rho} e^{-iku\hat{x}} \nonumber
\\
&=& \frac{1}{2} \int_{-1}^{+1} du \; W(u)
e^{iu\eta_x (\hat{b}_x+\hat{b}_x^\dagger )} \hat{\rho} 
e^{-iu\eta_x (\hat{b}_x+\hat{b}_x^\dagger )} \, .
\end{eqnarray}
Here, $\gamma$ is the spontaneous emission rate and 
$W(u)=(3/4)(1+u^2)$ describes the angular distribution of
spontaneous emission for an atomic dipole transition.

Staying in one dimension and considering the preparation of a
squeezed state of the motion, the appropriate combination of
laser fields is
\begin{eqnarray}
E_{\rm L}(\hat{x},t) &=& E_1(\hat{x},t)
+ E_2(\hat{x},t) \nonumber
\\
&=& \frac{{\cal E}}{\sqrt{2}}
\left[ e^{-i\eta_x(\hat{b}_x+\hat{b}_x^\dagger )} +
e^{-i\delta_{21}t+i\eta_x(\hat{b}_x+\hat{b}_x^\dagger )} \right] ,
\end{eqnarray}
with $\delta_{21}=2\nu_x$. 
Adiabatically eliminating the atomic excited state, which amounts to
setting
\begin{equation}
\hat{\sigma}_- \simeq i \frac{E_{\rm L}(\hat{x},t)}{\Delta_{01}}
\end{equation}
in (\ref{SpEme}), one finds that the leading order (in $\eta_x$) 
contribution to the motional dynamics arising from atomic recoil 
due to spontaneous emission takes the form
\begin{eqnarray}
s(t) &&
\left[ 2(\hat{b}_x+\hat{b}_x^\dagger )\hat{\rho} 
(\hat{b}_x+\hat{b}_x^\dagger )
- (\hat{b}_x+\hat{b}_x^\dagger )^2\hat{\rho} \right. \nonumber
\\
&& \;\;\; - \left.
\hat{\rho} (\hat{b}_x+\hat{b}_x^\dagger )^2 \right] ,
\end{eqnarray}
where
\begin{equation}
s(t) = \frac{1}{4}\frac{\gamma\eta_x^2{\cal E}^2}{\Delta_{01}^2} \left(
|1-e^{-i\delta_{21} t}|^2+\frac{1}{5}|1+e^{i\delta_{21}t}|^2 \right) .
\end{equation}
So, the rate at which atomic spontaneous emission can be expected to
influence the motional dynamics is on the order of 
$\gamma\eta_x^2{\cal E}^2/\Delta_{01}^2$, whereas the rate at which the
squeezed state is prepared is essentially 
$\chi = 4\eta_x^2{\cal E}^2/\Delta_{01}$, leading to the condition
\begin{equation}
\frac{\gamma}{\Delta_{01}} \ll 1
\end{equation}
for spontaneous emission to have negligible effect.

\subsection{
Motion-light coupling
}

The analysis of the effects of spontaneous emission in the context of
the atom-cavity state transfer scheme follows the above working, only
now the relevant substitution is
\begin{equation}
\hat{\sigma}_- \simeq - \frac{{\cal E}_{\rm A}(t)}{\Delta_{\rm 0A}} -
\frac{g_0}{\Delta_{\rm 0A}} \sin (k\hat{x})\hat{a} .
\end{equation}
Assuming that ${\cal E}_{\rm A}\gg\eta_x g_0\sqrt{\langle\hat{a}^\dagger
\hat{a}\rangle}$, to leading order in $\eta_x$ one obtains a term of
the form
\begin{eqnarray}
\eta_x^2 \, \frac{\gamma}{10} \frac{{\cal E}_{\rm A}^2}{\Delta_{\rm 0A}^2} 
&& \left[ 2(\hat{b}_x+\hat{b}_x^\dagger )\hat{\rho} 
(\hat{b}_x+\hat{b}_x^\dagger )
- (\hat{b}_x+\hat{b}_x^\dagger )^2\hat{\rho} \right. \nonumber
\\
&& \;\;\; - \left.
\hat{\rho} (\hat{b}_x+\hat{b}_x^\dagger )^2 \right] .
\end{eqnarray}
Hence, in order to be able to neglect the effects of spontaneous
emission on the transfer process, one requires that  
\begin{equation} \label{SpEcondition}
\Gamma = 
\frac{\eta_x^2g_0^2{\cal E}_{\rm A}^2}{\kappa\Delta_{\rm 0A}^2} \gg
\eta_x^2 \, \frac{\gamma}{10} \frac{{\cal E}_{\rm A}^2}{\Delta_{\rm 0A}^2}
\;\;\;\; {\rm or} \;\;\;\;
\frac{10g_0^2}{\kappa\gamma} \gg 1 \, .
\end{equation}
This, not surprisingly, corresponds to the regime of 
strong coupling in cavity QED.

\section{
Lamb-Dicke approximation for the atom-cavity coupling laser
}

In deriving our model for motion-light coupling we make the
(simplifying) assumption that position dependence of the external
coupling laser can be neglected, i.e., 
$E_{\rm A}(\hat{y},t)\simeq {\cal E}_{\rm A}(t)
e^{-i\phi_{\rm A}}$. We mentioned that this could be justified,
for instance, in the case where the laser field forms a standing
wave with the trap centered at an antinode, i.e., 
$E_{\rm A}(\hat{y},t)\propto\cos (k\hat{y})\simeq
1$, assuming tight confinement in the $y$-direction.

However, a traveling wave laser field (which is most likely a
simpler proposition from an experimental point of view) should also 
suffice, as the following argument shows.
Assume that 
\begin{equation}
E_{\rm A}(\hat{y},t) = {\cal E}_{\rm A}(t) e^{ik\hat{y}} .
\end{equation}
Then, first of all, 
$|E_{\rm A}(\hat{y},t)|^2={\cal E}_{\rm A}(t)^2$.
Secondly, on adiabatically eliminating the atomic excited state, we
have an interaction term of the form (omitting constant coefficients)
\begin{equation}
e^{i\eta_y(\hat{b}_y+\hat{b}_y^\dagger )} \sin [\eta_x 
(\hat{b}_x+\hat{b}_x^\dagger )](\hat{a}+\hat{a}^\dagger ) .
\end{equation}
To second order in the Lamb-Dicke parameters, this expression takes
the form
\begin{eqnarray} \label{xyexpans}
&& \eta_x (\hat{b}_x+\hat{b}_x^\dagger )(\hat{a}+\hat{a}^\dagger ) 
\nonumber
\\
&& \;\;\;\;\;\; + i\eta_x\eta_y (\hat{b}_y+\hat{b}_y^\dagger )
(\hat{b}_x+\hat{b}_x^\dagger )(\hat{a}+\hat{a}^\dagger ) .
\end{eqnarray}
Now, we have $\delta_{\rm cA}=\nu_x$, and assuming that $\nu_x$ and 
$\nu_y$ are both large,
and that $|2\nu_x-\nu_y|$ is also large, the only contribution in 
(\ref{xyexpans}) that is not rapidly rotating is the desired term
$\eta_x(\hat{a}^\dagger\hat{b}_x+\hat{b}_x^\dagger\hat{a})$. That is, 
the contribution from the position dependence of $E_{\rm A}(\hat{y},t)$,
as well as being of order $\eta_x\eta_y$, is also rapidly rotating
(assuming large $\nu_y$), and hence can be neglected.

\begin{table}
\caption{Two-mode squeezed state preparation: examples of fidelities
achievable for several different trapping configurations. Note that, in the
dimensionless units used here, the values of $\chi$ correspond to the choice
${\cal E}^2/\Delta_{01}=0.1$. 
Also, since $\eta_i\propto 1/\sqrt{\nu_i}$, setting
$\eta_x^\prime =\eta_z^\prime$ means that
$(\alpha /\beta )=\sqrt{\nu_x/\nu_z}$.
\label{table1}}
\begin{tabular}{cccccc}
$\eta_x^\prime ,\eta_z^\prime$ & $\nu_x$ & $\nu_z$ & $\chi$ & $r=\chi T$ & $F$\\ 
\tableline
0.1 & 1 & 3 & 0.004 & 1 & 0.991\\
0.1 & 1 & 3 & 0.004 & 1.5 & 0.932\\
0.1 & 1 & 4 & 0.004 & 1.5 & 0.955\\
0.0707 & 1 & 3 & 0.002 & 1 & 0.996\\
0.0707 & 1 & 3 & 0.002 & 1.5 & 0.975\\
0.0707 & 1 & 4 & 0.002 & 1.5 & 0.986\\
0.0577 & 1 & 3 & 0.00133 & 1 & 0.998\\
0.0577 & 1 & 3 & 0.00133 & 1.5 & 0.987\\
0.0577 & 1 & 4 & 0.00133 & 1.5 & 0.994\\
\end{tabular}
\end{table}

\begin{table}
\caption{Numerical results for transfer of the truncated phase state
$1/\sqrt{11}\sum_{n=0}^{10}|n\rangle_x$. Parameters are $\kappa =1$, 
$g_0^2/\Delta_{\rm 0A}=0.2$, $(g_0{\cal E}_{\rm A}^{\rm max})/
\Delta_{\rm 0A}=1.0$.
\label{table2}}
\begin{tabular}{ccc}
$\eta_x$ & $\nu_x(=\delta_{\rm cA})$ & 
$\langle\Psi (t_{\rm f})|\Psi (t_{\rm f})\rangle\;$ (no jump)\\ 
\tableline
0.1 & 5 & 0.65\\
0.1 & 10 & 0.90\\
0.1 & 20 & 0.96\\
0.0707 & 5 & 0.66\\
0.0707 & 10 & 0.91\\
0.0707 & 20 & 0.97\\
\end{tabular}
\end{table}

\begin{table}
\caption{Numerical results for transfer of the truncated phase state
$1/\sqrt{21}\sum_{n=0}^{20}|n\rangle_x$. Parameters are $\kappa =1$, 
$g_0^2/\Delta_{\rm 0A}=0.2$, $(g_0{\cal E}_{\rm A}^{\rm max})/
\Delta_{\rm 0A}=1.0$.
\label{table3}}
\begin{tabular}{ccc}
$\eta_x$ & $\nu_x(=\delta_{\rm cA})$ & 
$\langle\Psi (t_{\rm f})|\Psi (t_{\rm f})\rangle\;$ (no jump)\\ 
\tableline
0.1 & 10 & 0.79\\
0.1 & 20 & 0.88\\
0.0707 & 10 & 0.84\\
0.0707 & 20 & 0.94\\
\end{tabular}
\end{table}

\begin{table}
\caption{Numerical results for transfer of the Fock state
$|n=10\rangle_x$. Parameters are $\kappa =1$, 
$g_0^2/\Delta_{\rm 0A}=0.2$, $(g_0{\cal E}_{\rm A}^{\rm max})/
\Delta_{\rm 0A}=1.0$.
\label{table4}}
\begin{tabular}{ccc}
$\eta_x$ & $\nu_x(=\delta_{\rm cA})$ & 
$\langle\Psi (t_{\rm f})|\Psi (t_{\rm f})\rangle\;$ (no jump)\\ 
\tableline
0.1 & 10 & 0.82\\
0.1 & 20 & 0.92\\
0.0707 & 10 & 0.85\\
0.0707 & 20 & 0.95\\
\end{tabular}
\end{table}

\begin{table}
\caption{Numerical results for transfer of the Schr\"odinger Cat state
${\cal N}_+^{-1}(|\alpha\rangle_x+|-\alpha\rangle_x)$, with 
$\alpha =\sqrt{10}$.
Parameters are $\kappa =1$, 
$g_0^2/\Delta_{\rm 0A}=0.2$, $(g_0{\cal E}_{\rm A}^{\rm max})/
\Delta_{\rm 0A}=1.0$.
\label{table5}}
\begin{tabular}{ccc}
$\eta_x$ & $\nu_x(=\delta_{\rm cA})$ & 
$\langle\Psi (t_{\rm f})|\Psi (t_{\rm f})\rangle\;$ (no jump)\\ 
\tableline
0.1 & 10 & 0.81\\
0.1 & 20 & 0.91\\
0.0707 & 10 & 0.85\\
0.0707 & 20 & 0.95\\
\end{tabular}
\end{table}

%Figure1
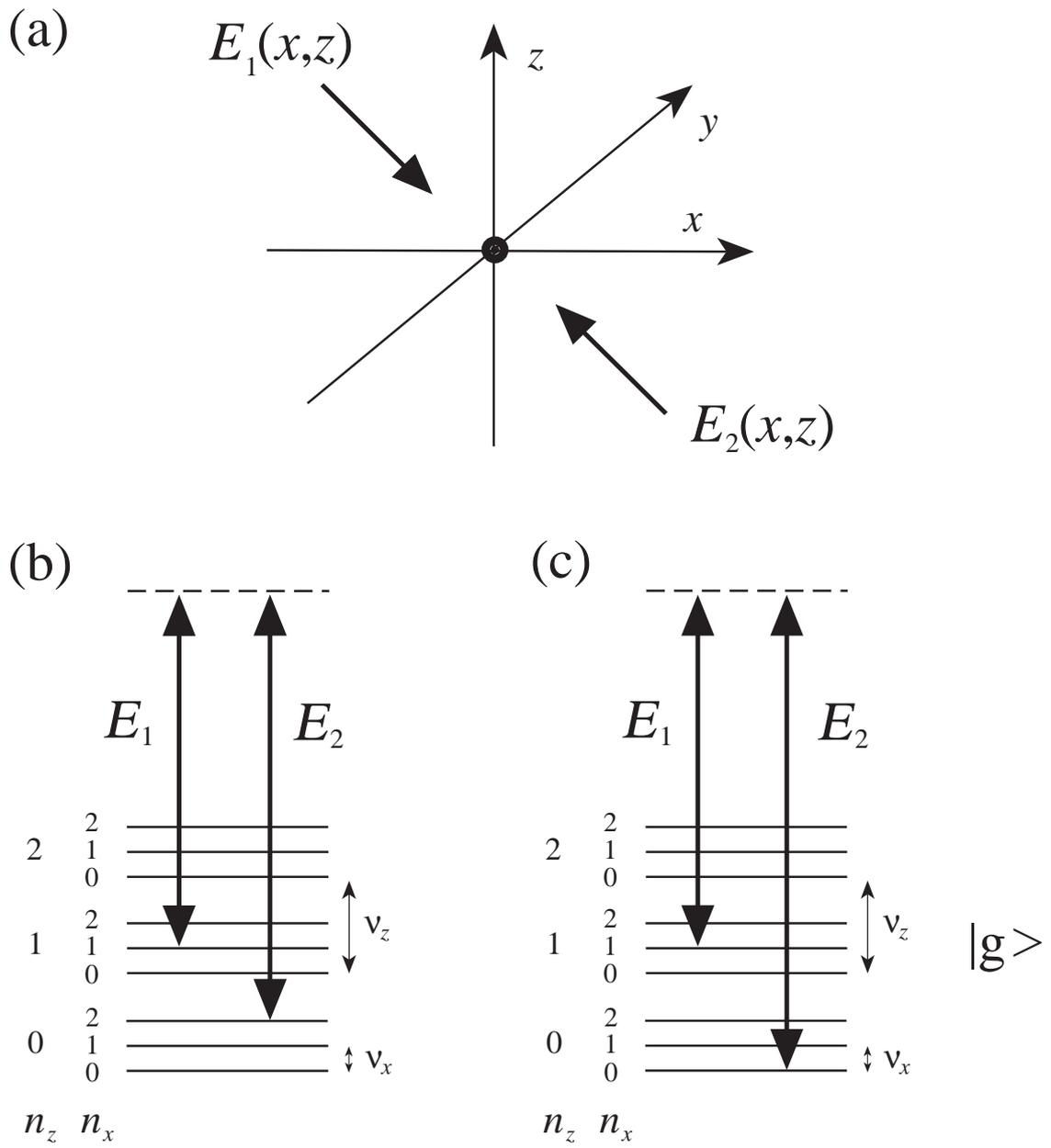
\begin{figure}
\caption{
(a) Coordinate axes and laser configuration for two-dimensional motional
state manipulation of a trapped atom. 
(The trapping potential is not shown.)
(b) Excitation scheme for the linear mixer interaction.
(c) Excitation scheme for preparation of a two-mode squeezed state of the
motion. Only the first few vibrational levels are shown, and the excited
atomic state is omitted for simplicity.
}
\end{figure}

%Figure2
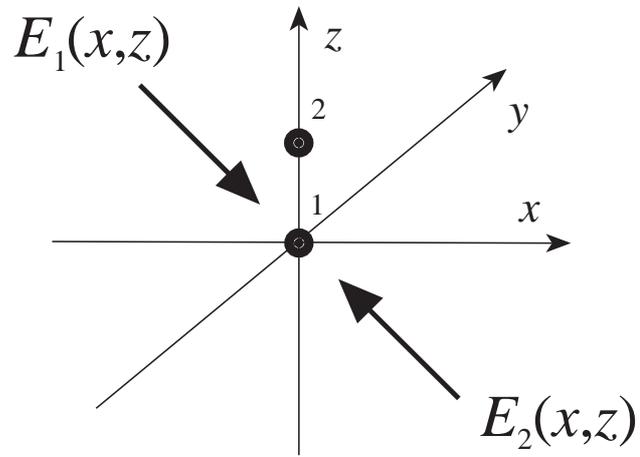
\begin{figure}
\caption{
Laser configuration for coupling two-ion collective and single-ion 
vibrational modes.
}
\end{figure}

%Figure3
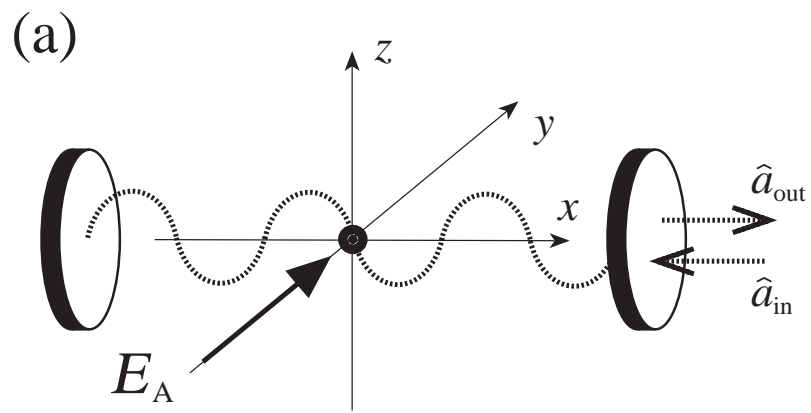
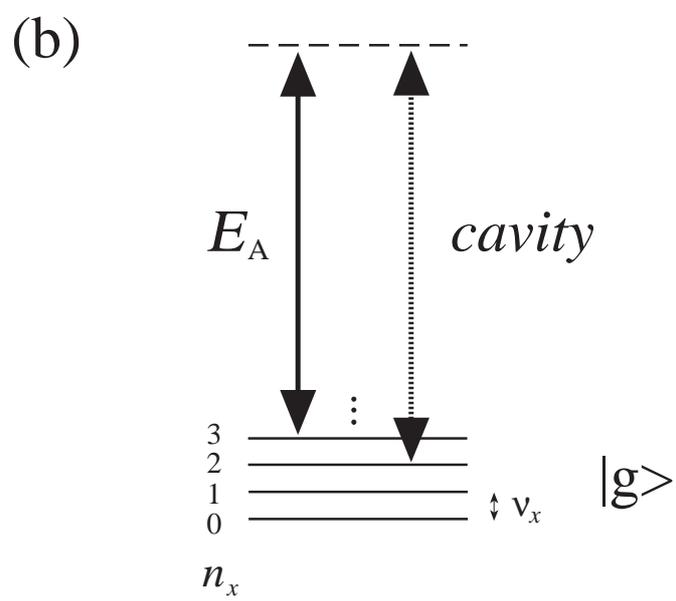
\begin{figure}
\caption{
Schematic of proposed (a) experimental setup and (b) excitation scheme for 
state transfer between the motion of a trapped atom or ion and a quantized
cavity mode of the electromagnetic field. All input and output to the 
cavity mode is through just one mirror, i.e., the other mirror is assumed 
to be perfect.
}
\end{figure}

%Figure4 
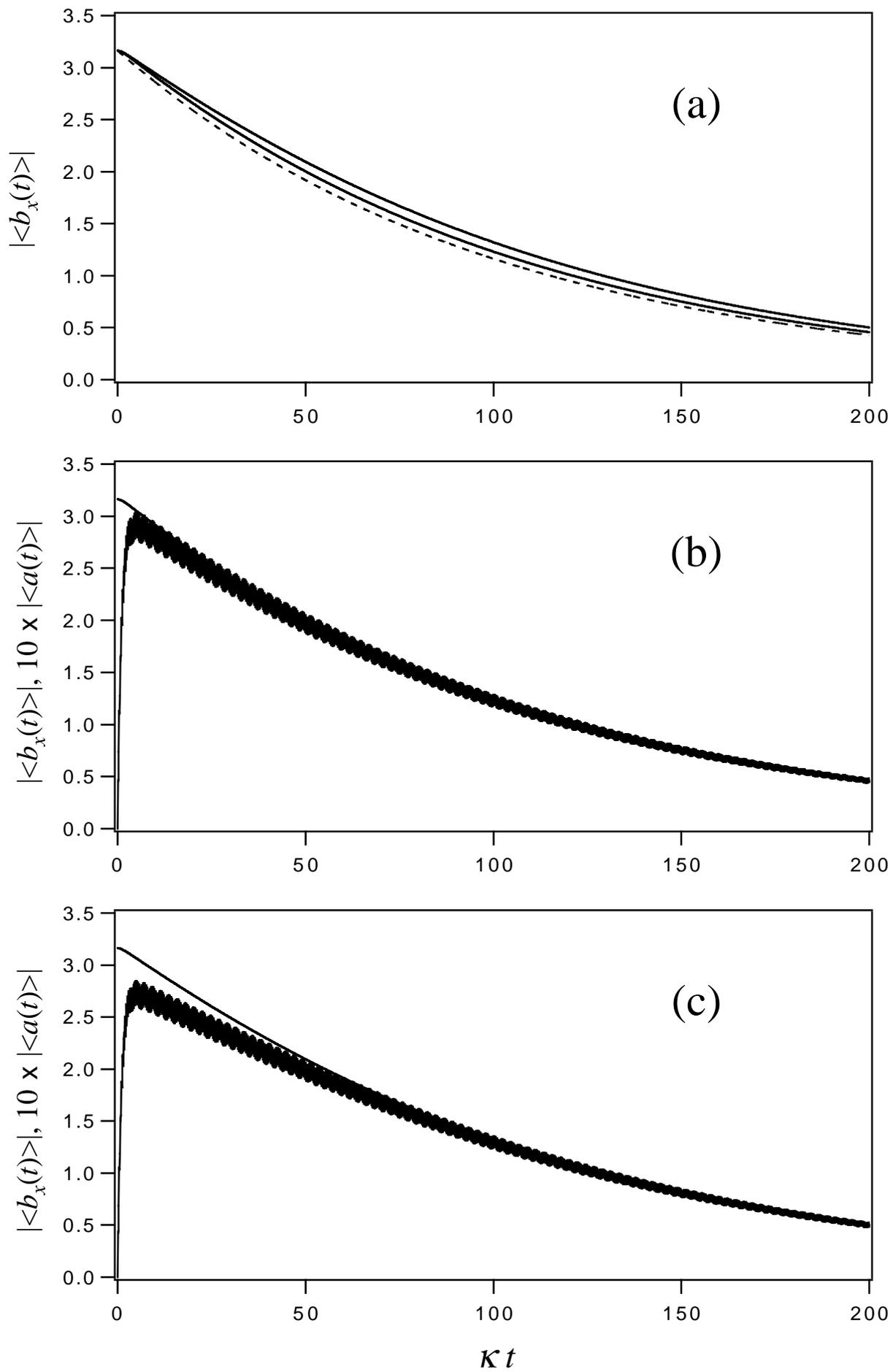
\begin{figure}
\caption{
(a) Decay of the motional mode amplitude $|\langle\hat{b}_x(t)\rangle |$ 
with time, given an initial coherent state of the motion 
$|\alpha =\sqrt{10}\,\rangle_x$, with $\eta_x=0.1$ (lower solid line), 
$0.15$ (upper solid line). The dashed curve is given by 
$\sqrt{10}\exp (-\Gamma t)$.
Other parameters are $\kappa =1$, $\nu_x=\delta_{\rm cA} =10$, 
$g_0^2/\Delta_{\rm 0A}=0.2$, and 
$\eta_xg_0{\cal E}_{\rm A}/\Delta_{\rm 0A}=0.1$ 
(corresponding to $\Gamma =0.01$). Figures (b) and (c) show the cavity 
field amplitude $|\langle\hat{a}(t)\rangle |$ ($\times 10$) and the
motional mode amplitude $|\langle\hat{b}_x(t)\rangle |$ for $\eta_x=0.1$
and $\eta_x=0.15$, respectively. Note that some of the structure (i.e., 
what looks like ``beats'') in $|\langle\hat{a}(t)\rangle |$ is due only 
to the finite number of points plotted.
}
\end{figure}

%Figure5
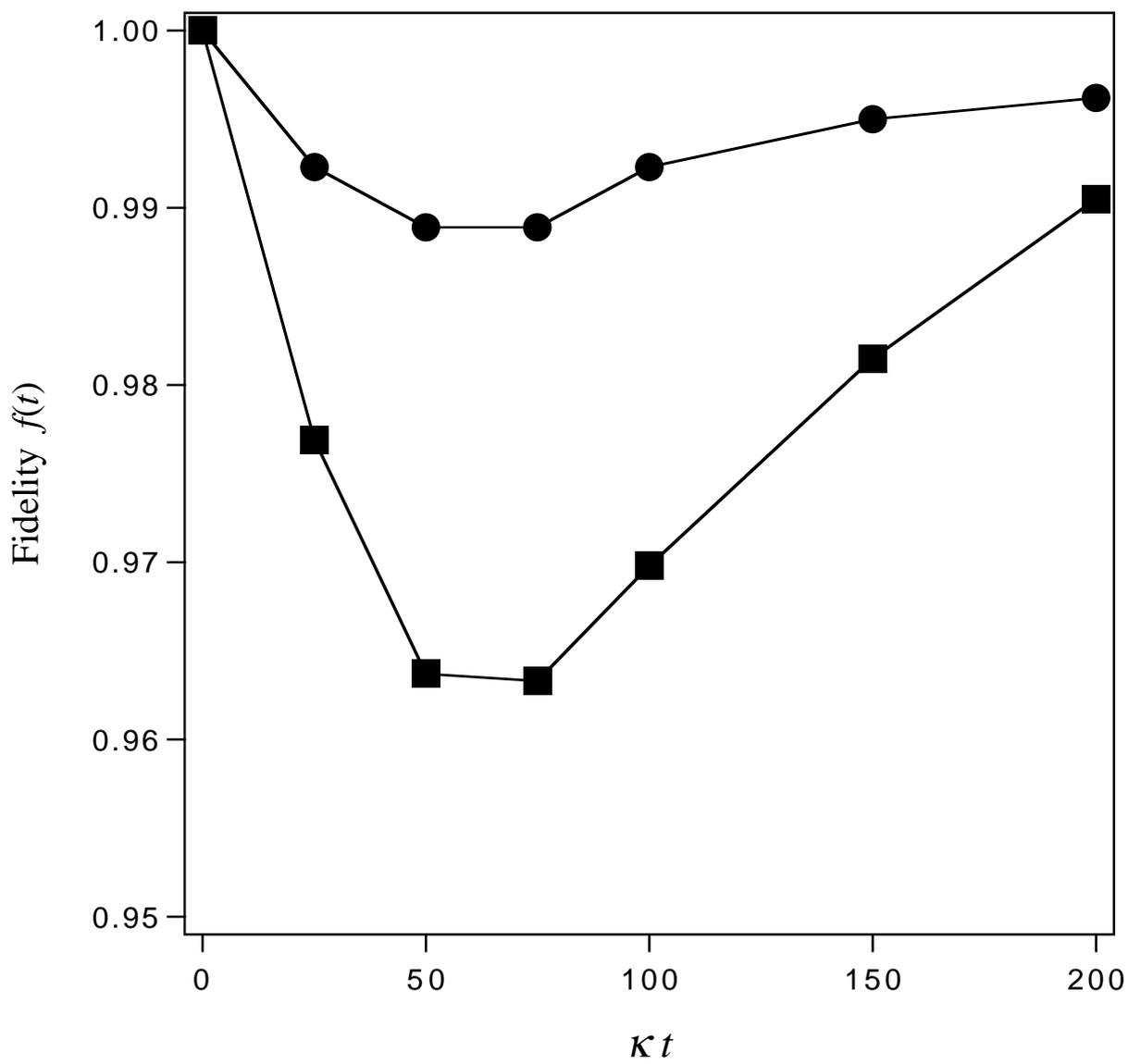
\begin{figure}
\caption{
Fidelity $f(t)$ as a function of time for the parameters of Figure 4 with 
$\eta_x=0.1$ (circles), and $\eta_x=0.15$ (squares).
}
\end{figure}

%Figure6
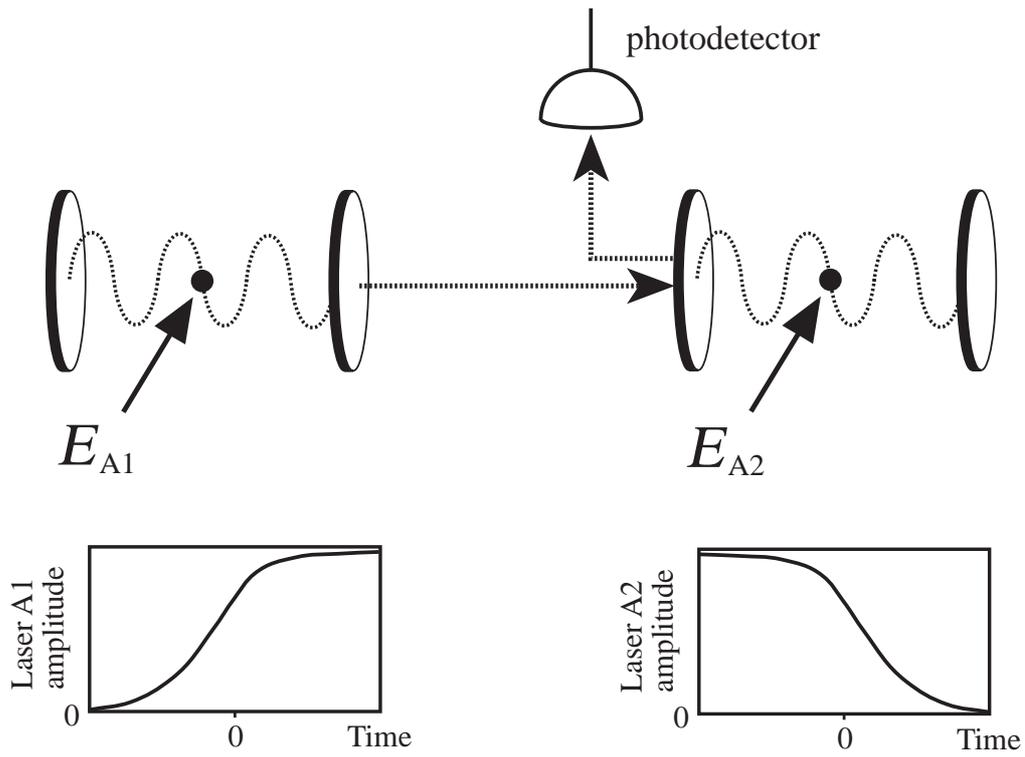
\begin{figure}
\caption{
Cascaded atom-cavity systems for motional state transfer. The coupling 
between cavities is assumed to be unidirectional (facilitated, for example, 
by Faraday isolators). During an ideal transfer no photons are detected by 
the photodetector, which monitors the output from the second cavity.
}
\end{figure}

%Figure7
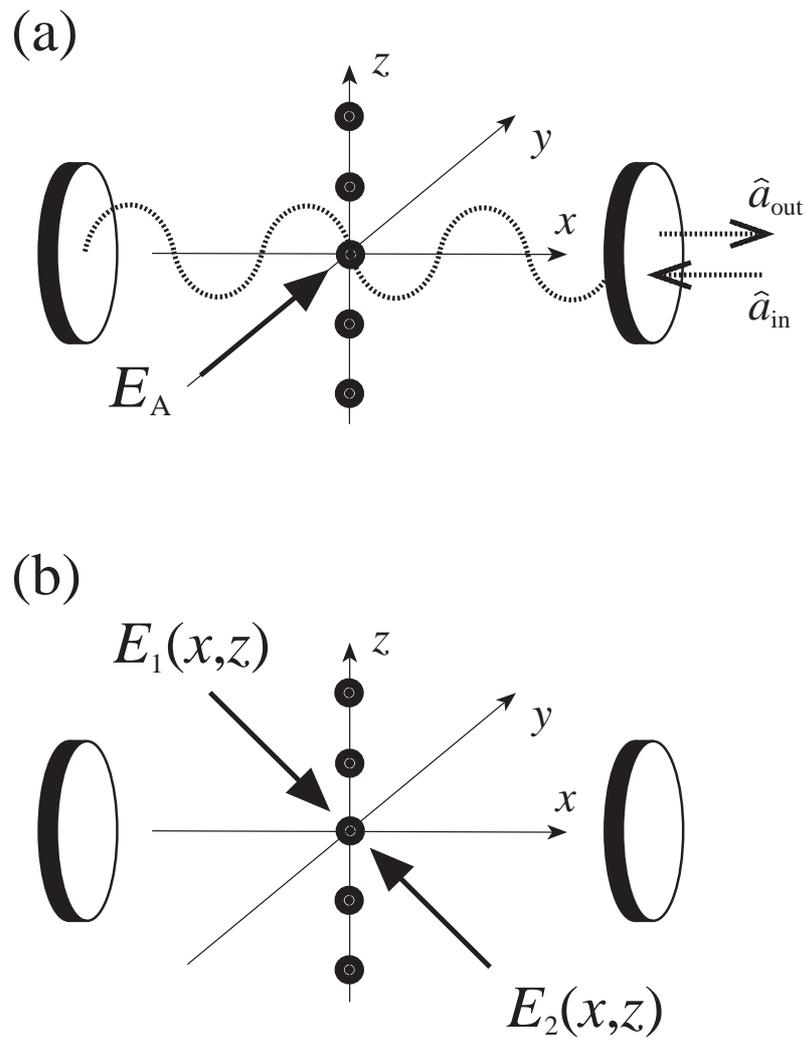
\begin{figure}
\caption{
Schematic for state transfer between light and collective motional modes
of a string of trapped ions.
(a) First, a single-ion interaction with the cavity field mode is used to
``receive'' an incoming quantum state, after which (b) auxiliary lasers
couple the single-ion and collective vibrational modes. 
}
\end{figure}


\begin{references}


\bibitem{Brune94}
M. Brune {\em et al}., Phys. Rev. Lett. {\bf72}, 3339 (1994);
G. Nogues {\em et al}., Nature {\bf400}, 239 (1999).

\bibitem{Turchette95}
Q.A. Turchette {\em et al}., Phys. Rev. Lett. {\bf75}, 4710 (1995).

\bibitem{Monroe95}
C. Monroe {\em et al}., Phys. Rev. Lett. {\bf75}, 4714 (1995).

\bibitem{Rauschenbeutel99}
A. Rauschenbeutel {\em et al}., Phys. Rev. Lett. {\bf83}, 5166 (1999).

\bibitem{Hagley97}
E. Hagley {\em et al}., Phys. Rev. Lett. {\bf79}, 1 (1997).

\bibitem{Turchette98}
Q.A. Turchette {\em et al}., Phys. Rev. Lett. {\bf81}, 3631 (1998).

\bibitem{Rauschenbeutel00}
A. Rauschenbeutel {\em et al}., Science {\bf288}, 2024 (2000).

\bibitem{Sackett00}
C.A. Sackett {\em et al}., Nature {\bf404}, 256 (2000).

\bibitem{Bouwmeester97}
D. Bouwmeester {\em et al}., Nature {\bf390}, 575 (1997).

\bibitem{Boschi98}
D. Boschi {\em et al}., Phys. Rev. Lett. {\bf80}, 1121 (1998).

\bibitem{Furusawa98}
A. Furusawa {\em et al.}., Science {\bf282}, 706 (1998).

\bibitem{Parkins99}
A.S. Parkins and H.J. Kimble, J. Opt. B:
Quantum Semiclass. Opt. {\bf1}, 496 (1999).

\bibitem{Parkins00a}
A.S. Parkins and H.J. Kimble, Phys. Rev. A {\bf61}, 052104 (2000).

\bibitem{Parkins00b}
A.S. Parkins and H.J. Kimble, in {\em Frontiers of Laser Physics and
Quantum Optics, Proceedings of the International Conference on Laser
Physics and Quantum Optics}, eds. Z. Xu, S. Xie, S.-Y. Zhu, and 
M.O. Scully (Springer, Berlin, 2000), p.321.
See also quant-ph/9909021.

\bibitem{Meekhof96}
D.M. Meekhof {\em et al}., Phys. Rev. Lett. {\bf76}, 1796 (1996).

\bibitem{Monroe96}
C. Monroe {\em et al}., Science {\bf272}, 1131 (1996).

\bibitem{Wineland98a}
D.J. Wineland {\em et al}., 
Jou. Res. Nat. Inst. Stand. Tech. {\bf103}, 259 (1998).

\bibitem{Roos99}
Ch. Roos {\em et al}., Phys. Rev. Lett. {\bf83}, 4713 (1999);
F. Schmidt-Kaler {\em et al}., quant-ph/0003096, submitted to
J. Mod. Opt..

\bibitem{Gou96a}
S.-C. Gou and P.L. Knight, Phys. Rev. A {\bf54}, 1682 (1996).

\bibitem{Gou96b}
S.-C. Gou, J. Steinbach, and P.L. Knight, Phys. Rev. A {\bf54}, 
R1014 (1996); {\em ibid}. {\bf54}, 4315 (1996).

\bibitem{Steinbach97}
J. Steinbach, J. Twamley, and P.L. Knight, Phys. Rev. A {\bf56},
4815 (1997).

\bibitem{Gardiner97}
S.A. Gardiner, J.I. Cirac, and P. Zoller, Phys. Rev. A {\bf55},
1683 (1997).

\bibitem{Drobny97}
G. Drobn\'y and B. Hladk\'y, Acta Phys. Slov. {\bf47}, 277 (1997).

\bibitem{Drobny98}
G. Drobn\'y, B. Hladk\'y, and V. Bu\u{z}ek, 
Phys. Rev. A {\bf58}, 2481 (1998).

\bibitem{Wineland98b}
D.J. Wineland {\em et al}., Physica Scripta {\bf T76}, 147 (1998).

\bibitem{Kneer98}
B. Kneer and C.K. Law, Phys. Rev. A {\bf57}, 2096 (1998).

\bibitem{Walls94}
D.F. Walls and G.J. Milburn, {\em Quantum Optics} 
(Springer-Verlag, Berlin, 1994).

\bibitem{Ou92b}
Z.Y. Ou, S.F. Pereira, and H.J. Kimble, Appl. Phys. B {\bf55},
265 (1992).

\bibitem{Einstein35}
A. Einstein, B. Podolsky, and N. Rosen, 
Phys. Rev. {\bf47}, 777 (1935).

\bibitem{Raizen92}
See, for example,
M.G. Raizen {\em et al}., Phys. Rev. A {\bf45}, 6493 (1992);
I. Waki {\em et al}., Phys. Rev. Lett. {\bf68}, 2007 (1992);
H.C. N\"agerl {\em et al}., Appl. Phys. B: Photophys. Laser Chem.
{\bf66}, 603 (1998);
R.J. Hughes {\em et al}., Fortschr. Phys. {\bf46}, 329 (1998).

\bibitem{Cirac95a}
J.I. Cirac and P. Zoller, Phys. Rev. Lett. {\bf74}, 4091 (1995).

\bibitem{James98}
D.F.V. James, Appl. Phys. B {\bf66}, 181 (1998).

\bibitem{Morigi99}
G. Morigi, J. Eschner, J.I. Cirac, and P. Zoller, 
Phys. Rev. A {\bf59}, 3797 (1999).

\bibitem{Zeng94}
H. Zeng and F. Lin, Phys. Rev. A {\bf 50}, R3589 (1994).

\bibitem{Harrison97}
Note that 
the terms of second order in $\eta_x$ correspond to a nonlinear
coupling between the modes which can, in a slightly different
configuration [in particular, with the coupling laser turned 
off, i.e., $\Omega (t)=0$, and with the cavity mode driven by a 
coherent field] be used for a quantum nondemolition
measurement of the {\em phonon number}; see
F.E. Harrison, A.S. Parkins, M.J. Collett, and D.F. Walls,
Phys. Rev. A {\bf55}, 4412 (1997).

\bibitem{Gardiner00}
C.W. Gardiner and P. Zoller, {\em Quantum Noise} (Springer-Verlag, 
Berlin, 2000).

\bibitem{Cirac97}
J.I. Cirac, P. Zoller, H.J. Kimble, and H. Mabuchi, 
Phys. Rev. Lett. {\bf78}, 3221 (1997).

\bibitem{Gardiner93}
C.W. Gardiner, Phys. Rev. Lett. {\bf70}, 2269 (1993).

\bibitem{Carmichael93}
H.J. Carmichael, Phys. Rev. Lett. {\bf70}, 2273 (1993).

\bibitem{Cochrane99}
P.T. Cochrane, G.J. Milburn, and W.J. Munro, Phys. Rev. A {\bf59}, 2631
(1999);
M.C. Oliveira and W.J. Munro, Phys. Rev. A {\bf61}, 042309 (2000).

\bibitem{Nambu00}
Y. Nambu {\em et al}., Phys. Rev. A {\bf62}, 012312 (2000).

\bibitem{Turchette00}
Q.A. Turchette {\em et al}., Phys. Rev. A {\bf61}, 063418 (2000).

\bibitem{Ye99}
J. Ye, D.W. Vernooy, and H.J. Kimble, Phys. Rev. Lett. {\bf83}, 4987
(1999).

\bibitem{Fortagh00}
J. Fortagh {\em et al}., Appl. Phys. B {\bf70}, 701 (2000);
D. Cassettari {\em et al}., Appl. Phys. B {\bf70}, 721 (2000).

\bibitem{Hood00}
C.J. Hood {\em et al}., Science {\bf287}, 1447 (2000).

\bibitem{Pinkse00}
P.W.H. Pinkse {\rm et al}., Nature {\bf404}, 365 (2000).

\bibitem{Rempe92}
G. Rempe {\em et al}., Opt. Lett. {\bf17}, 363 (1992).

\bibitem{Bcool}
Note that with a vacuum input to the cavity field, the dynamics 
described by Eq.(\ref{Bmodedamp}) will correspond simply to cooling of
the collective mode to its ground state. In this way, the collective mode
could be ``initialized''.

\bibitem{King98}
B.E. King {\em et al}., Phys. Rev. Lett. {\bf81}, 1525 (1998).
In this experiment with a pair of ${}^9{\rm Be}^+$ ions, the heating
rates of the modes of relative ion motion (for example, the ``stretch'' 
mode) were found to be significantly smaller than those of the 
center-of-mass modes

\bibitem{Vaidman94}
L. Vaidman, Phys. Rev. A {\bf49}, 1473 (1994). See also
L. Vaidman and N. Yoran, Phys. Rev. A {\bf59}, 116 (1999).

\bibitem{Braunstein98}
S.L. Braunstein and H.J. Kimble, Phys. Rev. Lett. {\bf80}, 869
(1998).

\bibitem{Parkins00c}
A.S. Parkins, quant-ph/0006113, submitted to J. Opt. B: Quantum Semiclass.
Opt.. In the proposal presented in this work, motion in only 
one dimension is considered, but auxiliary internal atomic states that are 
not coupled to the cavity field, and thereby do not participate in the
motional state transfer process, enable delocalized coherent or Fock 
states to be prepared.

\bibitem{Polzik99}
E.S. Polzik, Phys. Rev. A {\bf59}, 4202 (1999).

\bibitem{Lukin00}
M.D. Lukin, S.F. Yelin, and M. Fleischhauer, Phys. Rev. Lett. {\bf84},
4232 (2000).

\bibitem{Kozhekin00}
A.E. Kozhekin, K. M{\o}lmer, and E.S. Polzik, Phys. Rev. A {\bf62},
033809 (2000).

\bibitem{Duan00}
L.M. Duan {\em et al}., quant-ph/0003111.

\bibitem{Lloyd99}
S. Lloyd and S.L. Braunstein, Phys. Rev. Lett. {\bf82}, 1784 (1999).

\bibitem{Loock00}
P. van Loock and S.L. Braunstein, Phys. Rev. Lett. {\bf84}, 3482 (2000).

\bibitem{Braunstein00}
S.L. Braunstein and H.J. Kimble, Phys. Rev. A {\bf61}, 042302 (2000).

\bibitem{Ralph00}
T.C. Ralph, Phys. Rev. A {\bf61}, 010303(R) (2000).

\bibitem{Fry00}
E.S. Fry and T. Walther, Adv. At. Mol. Phys. {\bf42}, 1 (2000).

\bibitem{Cirac92}
See, for example, J.~I. Cirac {\em et al}., 
Phys. Rev. A {\bf46}, 2668 (1992); 
C. D'Helon and G.~J. Milburn, {\em ibid}. {\bf52}, 4755 (1995).

\end{references}
\end{document}